\documentclass{ar2e}

\usepackage{latexsym,bm,amsmath,amssymb,amsfonts}
\usepackage{epsfig,graphics}

 \newcommand{\rmd}{{\rm d}}   %ELS%

  \long\def\comment#1{ }
  \newcommand{\eqnum}[1]{eq.~\eqref{#1}}

  \newcommand{\abar}{\alpha_s}
  
  \newcommand{\del}{\partial}

  \newcommand{\mcal}{\mathcal}
  \newcommand{\rme}{{\rm e}}

  \newcommand{\Lam}{\Lambda_{_{\rm QCD}}}

  \newcommand{\order}[1]{\mcal{O}{(#1)}}

\def\empile#1\over#2{\mathrel{\mathop{\kern 0pt#1}\limits_{#2}}}
\newcommand{\slvarepsilon}{\raise.15ex\hbox{$/$}\kern-.53em\hbox{$\varepsilon$}}
\newcommand{\slL}{\raise.15ex\hbox{$/$}\kern-.53em\hbox{$L$}}
\newcommand{\slP}{\raise.15ex\hbox{$/$}\kern-.53em\hbox{$P$}}
\newcommand{\slp}{\raise.1ex\hbox{$/$}\kern-.63em\hbox{$p$}}
\newcommand{\slq}{\raise.1ex\hbox{$/$}\kern-.53em\hbox{$q$}}
\newcommand{\slv}{\raise.1ex\hbox{$/$}\kern-.63em\hbox{$v$}}
\newcommand{\slR}{\raise.15ex\hbox{$/$}\kern-.53em\hbox{$R$}}
\newcommand{\slQ}{\raise.15ex\hbox{$/$}\kern-.53em\hbox{$Q$}}
\newcommand{\slK}{\raise.15ex\hbox{$/$}\kern-.53em\hbox{$K$}}
\newcommand{\slk}{\raise.15ex\hbox{$/$}\kern-.53em\hbox{$k$}}
\newcommand{\slSigma}{\raise.15ex\hbox{$/$}\kern-.53em\hbox{$\Sigma$}}
\newcommand{\slcalP}{\raise.15ex\hbox{$/$}\kern-.63em\hbox{$\cal P$}}
\newcommand{\slA}{\raise.15ex\hbox{$/$}\kern-.73em\hbox{$A$}}
\newcommand{\slbfA}{\raise.15ex\hbox{$/$}\kern-.73em\hbox{${\imb A}$}}
\newcommand{\slpartial}{\raise.15ex\hbox{$/$}\kern-.53em\hbox{$\partial$}}
\newcommand{\sla}{\raise.15ex\hbox{$/$}\kern-.53em\hbox{$a$}}
\newcommand{\slb}{\raise.15ex\hbox{$/$}\kern-.53em\hbox{$b$}}
\newcommand{\slc}{\raise.15ex\hbox{$/$}\kern-.53em\hbox{$c$}}
\newcommand{\slC}{\raise.15ex\hbox{$/$}\kern-.63em\hbox{$C$}}

\def\p{{\boldsymbol p}}

\def\k{{\boldsymbol k}}

\def\x{{\boldsymbol x}}

\def\y{{\boldsymbol y}}

\def\r{{\boldsymbol r}}

\def\b{{\boldsymbol b}}

\def\bs{\boldsymbol}

 \def\simge{\mathrel{%
   \rlap{\raise 0.511ex \hbox{$>$}}{\lower 0.511ex \hbox{$\sim$}}}}
\def\simle{\mathrel{
   \rlap{\raise 0.511ex \hbox{$<$}}{\lower 0.511ex \hbox{$\sim$}}}}

\begin{document}
%\input epsf.tex    %<-If you need EPS figures to be
                   %  called in {figure} environment for PC
%\input epsf.def   %<-If you need EPS figures to be
                   %  called in {figure} environment for Macintosh

\jname{Annu. Rev. Nucl. Part. Sci.} 
\jyear{}
\jvol{}
\ARinfo{}

\title{The Color Glass Condensate}

\markboth{Fran\c cois Gelis et al.}{The Color Glass Condensate}

\author{
Fran\c cois Gelis, Edmond Iancu
\affiliation{IPhT, CEA/Saclay, 91191 Gif sur Yvette cedex, France\\
             francois.gelis@cea.fr, edmond.iancu@cea.fr}
Jamal Jalilian-Marian
\affiliation{Department of Natural Sciences, 
Baruch College, 17 Lexington Ave\\ New York, NY-10010, USA\\
CUNY Graduate Center, 365 fifth Ave, New York, NY-10016, USA
\\
jamal.jalilian-marian@baruch.cuny.edu}
Raju Venugopalan
\affiliation{Physics Department, BNL, Upton NY-11973, USA\\
raju@bnl.gov}
}


\begin{abstract}
  \baselineskip 16pt We provide a broad overview of the theoretical
  status and phenomenological applications of the Color Glass
  Condensate effective field theory describing universal properties of
  saturated gluons in hadron wavefunctions that are extracted from
  deeply inelastic scattering and hadron-hadron collision experiments
  at high energies.

%and the
%  interactions of saturated gluons in QCD.
\end{abstract}

\maketitle
\topmargin 2cm

\baselineskip 18pt 
\newpage

\section{High energy QCD and the Color Glass Condensate}

Only 5\% of the mass of the universe is ``bright matter'', but 99\% of
this visible matter is described by QCD, the theory of the strong
interactions.  QCD has been described as a nearly perfect theory, with
the only parameters being the masses of the current
quarks~\cite{Wilcz1}. The vast bulk of visible matter is therefore
``emergent'' phenomena arising from the rich dynamics of the QCD
vacuum and the interactions of the fundamental quark and gluon
constituents of QCD. Although we have enough information from a wide
range of experimental data and from numerical lattice computations
that QCD is the right theory of the strong interactions, outstanding
questions remain about how a variety of striking phenomena arise in
the theory.  This is because the complexity of the theory also makes
it very difficult to solve.

Our focus in this review is on high energy scattering in QCD. A
traditional approach to these phenomena is to divide them into ``hard''
or ``soft'' scattering, corresponding respectively to large or small
momentum exchanges in the scattering. In the former case, because of
the ``asymptotic freedom'' of QCD, phenomena such as jets can be
computed in a perturbative framework. In the latter case, because the
momentum transfer is small, the ``infrared slavery'' of QCD suggests
that the coupling is large; the scattering therefore is intrinsically
non-per\-tur\-ba\-ti\-ve and therefore not amenable to first principles
analysis. This is problematic because ``soft'' dynamics comprises the
bulk of QCD cross-sections. In contrast, the perturbatively calculable
hard cross-sections are rare processes\footnote{Lattice QCD is not of
  much help here because it is best suited to compute static
  properties of the theory.}.

We shall argue here that the traditional separation of hard versus
soft QCD dynamics is oversimplified because novel semi-hard scales
generated dynamically at high energies allow one to understand highly
non-perturbative phenomena in QCD using weak coupling methods.  To
clarify what we mean, consider inclusive cross-sections in deeply
inelastic scattering (DIS) experiments of electrons off hadrons.

\begin{figure*}[htb!]
  \begin{center}
    \resizebox*{5cm}{!}{\includegraphics{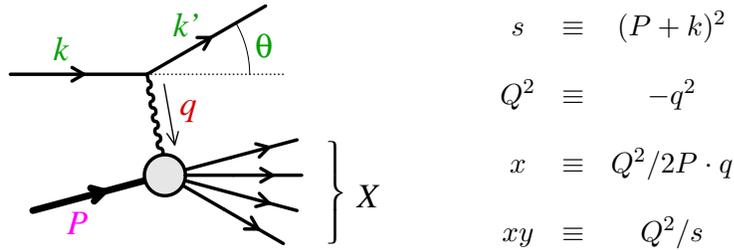}}
    \hskip 1cm
    \raise 16mm\hbox to 5cm{
      $\displaystyle{\begin{matrix}
          & s&\equiv& (P+k)^2\cr
          &Q^2&\equiv& -q^2\cr
          &x&\equiv&Q^2 / 2P\cdot q\cr
          &xy&\equiv& Q^2/s\cr
        \end{matrix}}$\hfil
    }
  \end{center}
  \caption{\baselineskip 15pt  \label{fig:DISkinematics}\sl Kinematics of Deep Inelastic
    Scattering.}
\end{figure*}
Inclusive cross-sections in DIS (see figure
\ref{fig:DISkinematics}) can be expressed in terms of the Lorentz
invariants i) $x$ which corresponds, at lowest order in perturbation
theory, to the longitudinal momentum fraction carried by a parton in
the hadron, ii) the virtual photon four-momentum squared $q^2 = -Q^2 <0$
exchanged between the electron and the hadron, iii) the inelasticity
$y$, the ratio of the photon energy to the electron energy in the
hadron rest frame, and iv) the center of mass energy squared $s$.
These satisfy the relation $x = Q^2/s y$.  Instead of the dichotomy of
``hard'' versus ``soft'', it is useful, for fixed $y$, to consider two asymptotic
limits in DIS that better illustrate the QCD dynamics of high energy
hadron wavefunctions. The first, called the Bjorken limit, corresponds
to fixed $x$ with $Q^2, s\rightarrow \infty$. The second is the
Regge-Gribov limit of fixed $Q^2$, $x\rightarrow 0$ and $s\rightarrow
\infty$.

\begin{figure*}[htb!]
\begin{center}
 \includegraphics[width=0.43\textwidth]{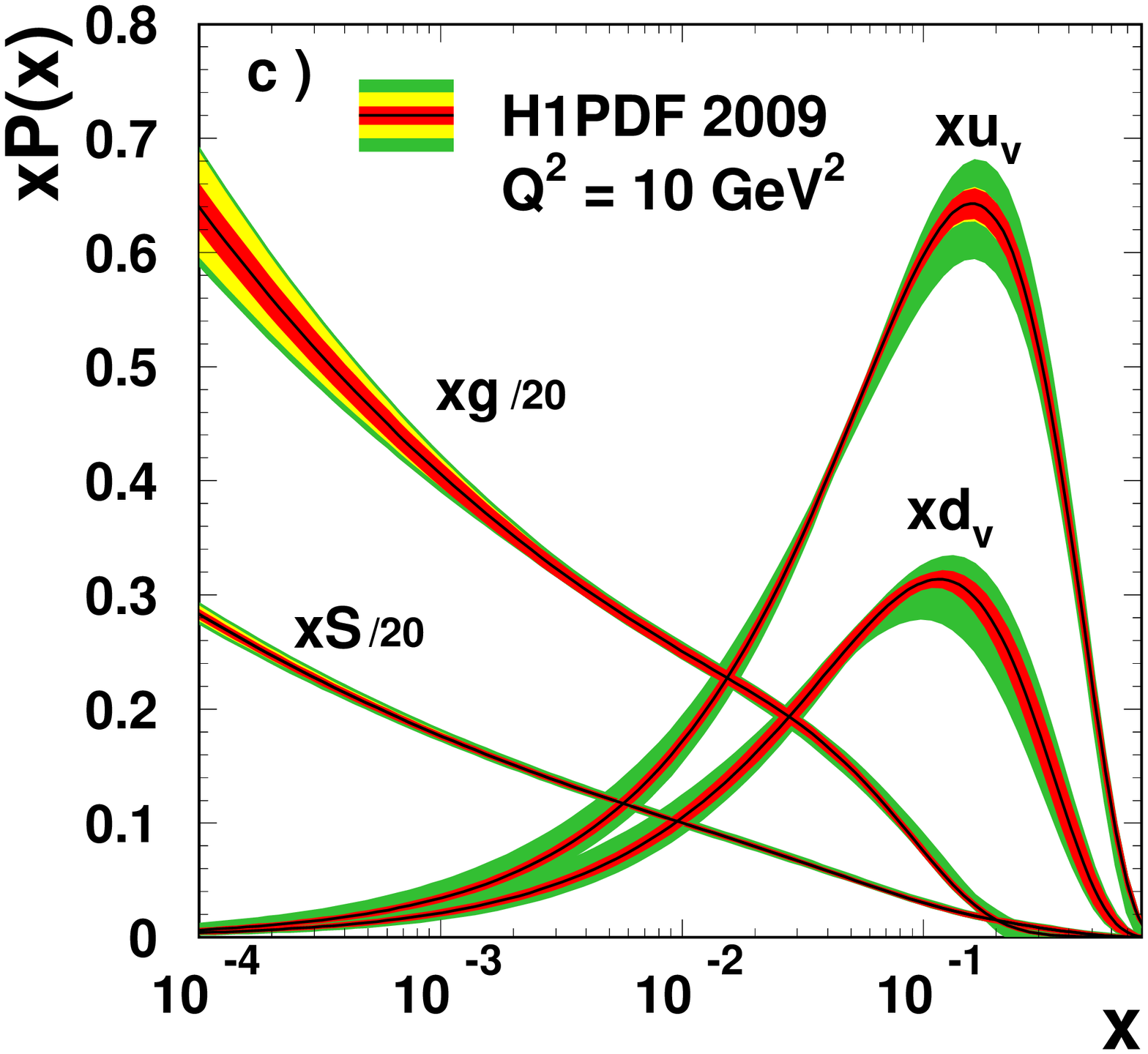}
 \hskip 3mm
 \includegraphics[width=0.45\textwidth]{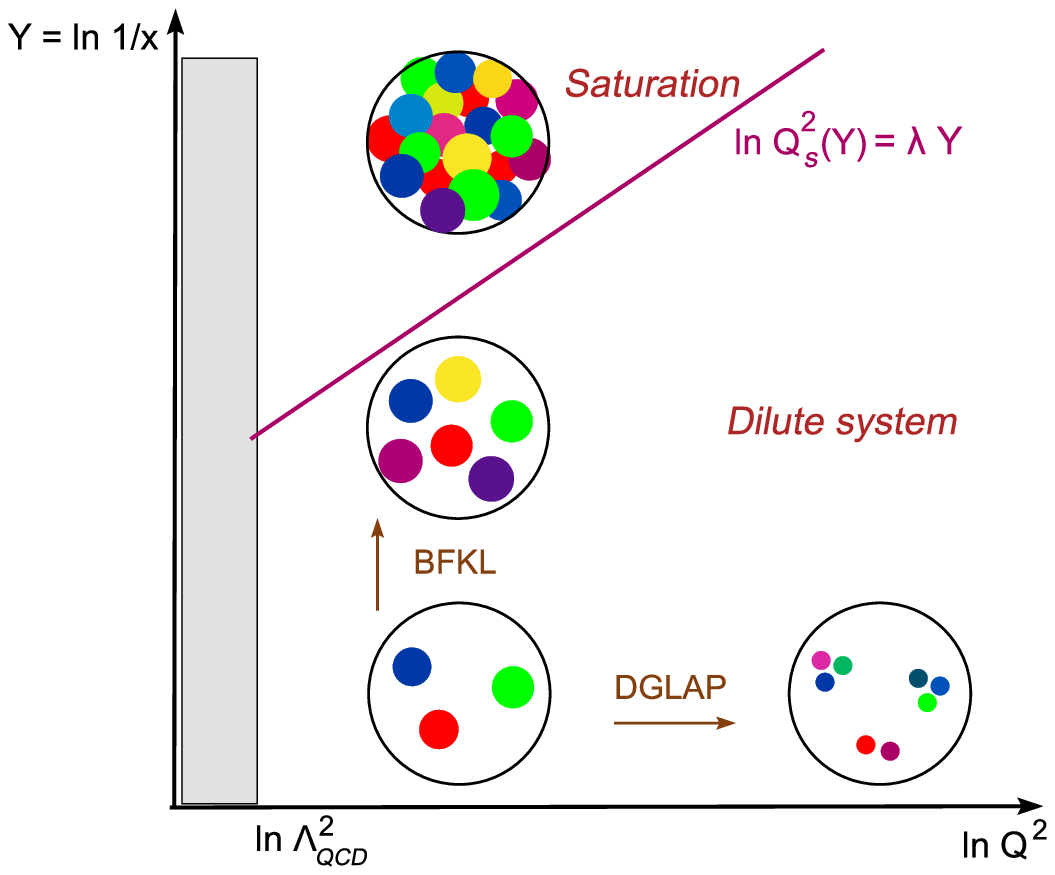}
\end{center}
\caption{\baselineskip 15pt  \label{fig:HERA-gluon}\sl Left: the
  $x$-evolution of the gluon, sea quark, and valence quark
  distributions for $Q^2=10$ GeV$^2$ measured at HERA~\cite{Aarona1}.
  Right: the ``phase--diagram'' for QCD evolution; each colored dot
  represents a parton with transverse area $\delta S_\perp \sim 1/Q^2$
  and longitudinal momentum $k^+=x P^+$. }
\end{figure*}

In the Bjorken limit of QCD, one obtains the parton model wherein the
hadron is viewed, in the infinite momentum frame (IMF), as a dilute
collection of valence quarks and ``wee'' (in the terminology coined
by Feynman) partons--small $x$ gluons and sea quark pairs. Albeit the number of
wee partons is large at high energies, the hadron is dilute (as illustrated in
figure~\ref{fig:HERA-gluon}) because the phase space density is very
small for $Q^2\rightarrow \infty$. In this limit, to leading order in
the coupling constant, the interaction of a probe with the hadron, on
the characteristic time scale $1/Q$, can be expressed as a hard
interaction with an individual parton. In this ``impulse
approximation'', the interaction of the struck parton in the hadron
with co--moving partons is suppressed due to time dilation. The
separation of hard and soft scattering alluded to previously is valid
here. Powerful tools such as the Operator Product Expansion (OPE) and
factorization theorems extend, to high orders in the coupling constant
expansion, the hard-soft separation between process-dependent physics
at the scale $1/Q$ and universal, long distance, non-perturbative Parton
Distribution Functions (PDFs). The evolution of the separation of hard
and soft scales is given by renormalization group (RG) equations
called the DGLAP (Dokshitzer-Gribov-Lipatov-Altarelli-Parisi)
equations~\cite{DGLAP}.

The study of strong interactions in the Regge-Gribov limit predates
QCD, and underlies concepts such as soft Pomeron and Reggeon
exchanges, whose dynamics, described by ``Reggeon Field Theory''
(RFT), was believed to encompass much of the phenomena of
multi-particle production. Albeit phenomenologically very
suggestive~\cite{DonnaL1}, the dynamics of RFT is intrinsically
non-perturbative and therefore not easily amenable to systematic
computations.  With the advent of high energy colliders, hadron
structure in the Regge-Gribov limit can be explored with $Q^2\geq 1$
GeV$^2$. As strikingly demonstrated by the HERA DIS data shown in
figure~\ref{fig:HERA-gluon}, the gluon distribution $x G(x,Q^2)$ in a
proton rises very fast with decreasing $x$ at large, fixed $Q^2$ --
roughly, as a power $x^{-\lambda}$ with $\lambda\simeq 0.3$. In the
IMF frame of the parton model, $x G(x,Q^2)$ is the number of gluons
with a transverse area $\delta S_\perp \ge 1/Q^2$ and a fraction
$k^+/P^+ \sim x$ of the proton longitudinal momentum\footnote{The
  light cone co-ordinates are defined as $k^\pm = (k^0 \pm
  k^3)/\sqrt{2}$.}.  In the Regge-Gribov limit, the rapid rise of the
gluon distribution at small $x$ is given by the BFKL
(Balitsky-Fadin-Kuraev-Lipatov) equation~\cite{BFKL}, which we will
discuss at some length later.

The stability of the theory formulated in the IMF requires that gluons
have a maximal occupation number of order $1/\alpha_s$. This bound is
saturated for gluon modes with transverse momenta $k_\perp \leq Q_s$,
where $Q_s(x)$ is a semi-hard scale, the ``saturation scale'', that
grows as $x$ decreases.  In this novel ``saturation'' regime of
QCD~\cite{saturation}, illustrated in figure~\ref{fig:HERA-gluon}
(right panel), the proton becomes a dense many body system of gluons.
In addition to the strong $x$ dependence, the saturation scale $Q_s$
has an $A$ dependence because of the Lorentz contraction of the
nuclear parton density in the probe rest frame.  The dynamics of
gluons in the saturation regime is non-perturbative as is typical of
strongly correlated systems. However, in a fundamental departure from
RFT, this dynamics can be computed using weak coupling methods as a
consequence of the large saturation scale dynamically generated by
gluon interactions. Thus instead of the ``hard'' plus ``soft''
paradigm of the Bjorken limit, one has a powerful new paradigm in the
Regge-Gribov limit to compute the bulk of previously considered
intractable scattering dynamics in hadrons and nuclei.

The Color Glass Condensate (CGC) is the description of the properties
of saturated gluons in the IMF in the Regge-Gribov limit. The
effective degrees of freedom in this framework are color sources
$\rho$ at large $x$ and gauge fields $A^\mu$ at small $x$. At high
energies, because of time dilation, the former are frozen color 
configurations on the natural time scales of the strong interactions
and are distributed randomly from event to event. The latter are
dynamical fields coupled to the static color sources. It is the
stochastic nature of the sources, combined with the separation of time
scales, that justify the ``glass'' appellation. The ``condensate''
designation comes from the fact that in the IMF, saturated gluons have
large occupation numbers ${\cal O}(1/\alpha_s)$, with typical momenta
peaked about a characteristic value $k_\perp \sim Q_s$. The dynamical
features of the CGC are captured by the JIMWLK\footnote{JIMWLK
  $\equiv$ Jalilian-Marian, Iancu, McLerran, Weigert, Leonidov,
  Kovner.} renormalization group (RG) equation that describes how the
statistical distribution $W[\rho]$ of the fast sources at a given $x$
scale evolves with decreasing $x$.  The JIMWLK RG is Wilsonian in
nature because weakly coupled fields integrated out at a given step in
the evolution are interpreted as ``induced color charges'' that modify
the statistical weight distribution.

The CGC framework is quite powerful because, given an initial
non-per\-tur\-ba\-ti\-ve distribution of sources at an initial scale
$x_0$, it allows one to compute systematically $n$-point gluon correlation functions
and their evolution with $x$ order by order in perturbation theory. In analogy to parton distribution
functions in the Bjorken limit, the distribution $W[\rho]$ captures
the properties of saturated gluons.  Unlike PDFs however, which in the
IMF are parton densities, $W[\rho]$ carries much more information.
Further, in contrast to the ``twist'' expansion of the Bjorken limit
which becomes extremely cumbersome beyond the leading order in $1/Q$,
the CGC framework includes all twist correlations. Like the
PDFs, the $W$'s are universal: the same distributions appear in
computations of inclusive quantities in both lepton-nucleus and
hadron-nucleus collisions.

The CGC is interesting in its own right in what it reveals about the
collective dynamics of QCD at high parton densities. The CGC RG
equations indicate that --at fixed impact parameter-- a proton and a
heavy nucleus become indistinguishable at high energy; the physics of
saturated gluons is universal and independent of the details of the
fragmentation region.  This universal dynamics has a correspondence
with reaction--diffusion processes in statistical physics. In
particular, it may lie in a ``spin glass'' universality class.
Understanding the nature of color screening and ``long range order'' in
this universal dynamics offers possibilities for progress in resolving
fundamental QCD questions regarding properties such as confinement and
chiral symmetry. A specific area of progress is in the mapping of the
CGC degrees of freedom to the traditional language of Pomerons and the
consequent prospects for understanding soft QCD dynamics. In nuclear
collisions, CGC dynamics produces ``Glasma'' field configurations at
early times: strong longitudinal chromo-electric and chromo-magnetic
fields color screened on transverse distance scales $1/Q_s$.  These generate
long range rapidity correlations, ``sphaleron-like'' topological
fluctuations characterized by large Chern-Simons charge, and
instabilities analogous to those seen in QED plasmas.

The CGC framework is applicable to a wide variety of processes in
e+p/A, p+A and A+A collisions. It provides an {\it ab initio} approach
to study thermalization in heavy ion collisions and the initial
conditions for the evolution of a thermalized Quark Gluon Plasma
(QGP). The interaction of hard probes with the Glasma is little
understood and is important for quantifying the transport properties
of the QGP precursor. A further phenomenological application of the
CGC is to the physics of ultra high energy cosmic rays \cite{cosmic}.

There are several reviews that discuss various aspects of the CGC and
its applications in depth~\cite{reviews1,reviews2,GelisLV2}.  Our aim
here is to provide a broad pedagogical overview of the current status
of theory and phenomenology for non-experts. The next section will
focus on the theoretical status, while section 3 discusses the CGC in
the context of DIS, hadronic and nuclear collisions. Section 4
presents the experimental results that probe the dynamics of saturated
gluons. We will end this review with a brief outlook.
  
\section{Color Glass Condensate: theoretical status}
We will begin this section with an elementary discussion of QCD bremsstrahlung in the
Regge-Gribov limit and use this discussion to motivate the BFKL
equation, how it leads to gluon saturation and describe key
features of saturation. The theoretical status of the CGC effective theory is then
presented. The section ends with a discussion of some advanced
topics that highlight open issues.

\subsection{Parton evolution at high energy}
\label{sec:BFKL}
\begin{figure*}[htb!]
\begin{center}
 \includegraphics[width=0.70\textwidth]{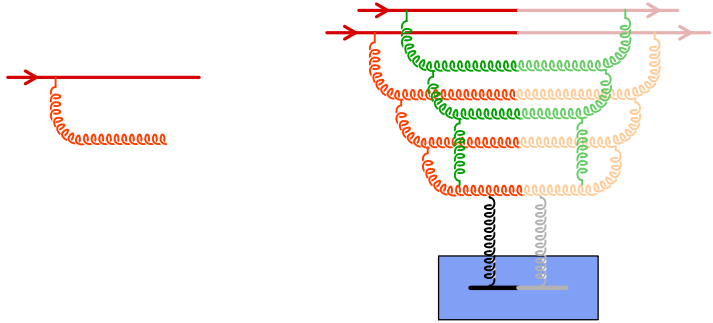}
\end{center}
\caption{\baselineskip 15pt  \label{fig:cascade}\sl Left: elementary bremsstrahlung
  radiation. Right: high energy scattering with evolution and
  recombination. }
\end{figure*}

The structure of a hadron depends upon the scales resolved by an
external probe. Quantum fields bound within the hadron can radiate
other virtual quanta which, by the uncertainty principle, live only
for a short period of time. Therefore, what looks like an isolated
quantum at a given resolution scale reveals a complicated substructure
when probed with finer space--time resolution. The structure of the
hadron in the IMF is specified by the longitudinal ($k^+$) and
transverse ($k_\perp$) momentum distributions of its quanta and by
their correlations--the latter become increasingly important with
increasing density.

In perturbative QCD, parton evolution proceeds via bremsstrahlung,
which favors the emission of {\em soft} and {\em collinear} gluons.
The left part of figure~\ref{fig:cascade} illustrates this elementary
radiation process.  When $x\ll 1$, to lowest order in $\alpha_s$, the
differential probability for this emission is given by
\begin{eqnarray}\label{brem} \rmd P_{\rm
    Brem}\,\simeq\,\frac{\alpha_s
    C_R}{\pi^2}\,\frac{\rmd^2k_\perp}{k_\perp^2}\,\frac{\rmd
    x}{x}\; ,
\end{eqnarray} 
where $C_R$ is the SU$(N_c)$ Casimir in the color representation of
the emitter-- $N_c$ for a gluon and $(N_c^2-1)/2N_c$ for a quark.
This formula demonstrates the collinear ($k_\perp\to 0$) and soft
($x\to 0$) singularities mentioned above, which produce a logarithmic
enhancement of gluon emission at small $k_\perp$ and/or $x$.  If the
emitter at small $x$ were a quark instead a gluon, there would be no
small $x$ enhancement; while the collinear enhancement is present for
either emitter. This asymmetry is due to the spin $1$ nature of the
gluon.

A high energy scattering such as the one illustrated in the right part
of figure~\ref{fig:cascade} probes the number of gluons with a given
$x$ and transverse momenta $k_\perp\le Q$. From eq.~(\ref{brem}), the
gluon number is
\begin{eqnarray}\label{xGxp}
x \frac{\rmd N_g}{\rmd x}(Q^2)\,=\,\frac{\alpha_s
 C_R}{\pi}\,\int_{\Lam^2}^{Q^2}\frac{\rmd k_\perp^2}{k_\perp^2}
 \,=\,\frac{\alpha_s C_R}{\pi}\,\ln\left(\frac{Q^2}{\Lam^2}\right)\; ,
\label{eq:Ng-1}
\end{eqnarray}
where the cutoff $\Lam$ has been introduced as a crude way to account
for confinement: when confined inside a hadron, a parton has a minimum
virtuality of $\order{\Lam^2}$.

In QCD, gluons can radiate softer gluons and thus rapidly multiply as
illustrated in figure~\ref{fig:cascade}.  Each subsequent emission is
$\alpha_s$ suppressed however when the final value of $x$ is small, these
corrections become large. For a cascade with $n$ intermediate gluons
strongly ordered in $x$, one obtains
 \begin{eqnarray}
 \alpha_s^n\,\int_{x}^1\frac{\rmd x_n}{x_n}
 \int_{x_n}^1\frac{\rmd x_{n-1}}{x_{n-1}}\cdots
 \int_{x_2}^1\frac{\rmd x_1}{x_1}
 \,=\,\frac{1}{n!}\left(\abar\ln\frac{1}{x}\right)^n
\; .
 \end{eqnarray}
 When $\alpha_s\ln(1/x)\gtrsim 1$, the correct result for the gluon
 distribution is obtained by summing contributions from all such
 ladders. The sum {\em exponentiates}, modifying eq.~(\ref{eq:Ng-1})
 into
 \begin{eqnarray}\label{eq:unintp}
x \frac{\rmd N_g}{\rmd x \rmd k_\perp^2}\,\sim\,\frac{\alpha_s
 C_R}{\pi}\,\frac{1}{k_\perp^2}\,\rme^{\omega\alpha_s Y}\; ,\qquad
 Y\equiv\ln \frac{1}{x}\;,
 \end{eqnarray}
 where $\omega$ is a number of order unity which is not fixed by this
 rough estimate. The variable $Y$ is known as the {\em rapidity}.

 To go beyond this simple power counting argument, one must treat more
 accurately the kinematics of the ladder diagrams and include the
 associated virtual corrections. The result is the previously
 mentioned {\em BFKL equation} for the $Y$-evolution of the
 unintegrated gluon distribution. The solution of this equation, which
 resums perturbative corrections $(\alpha_s Y)^n$ to all orders,
 confirms the exponential increase in eq.~(\ref{eq:unintp}), albeit
 with a $k_\perp$-dependent exponent and modifications to the
 $k_\perp^{-2}$-spectrum of the emitted gluons.

 An important property of the BFKL ladder is {\em coherence in time}.
 Because the lifetime of a parton in the IMF, $\Delta x^+\sim
 k^+/k_\perp^2\propto x$, the ``slow'' gluons at the lower end of the
 cascade have a much shorter lifetime than the preceding ``fast''
 gluons. Therefore, for the purposes of small $x$ dynamics, fast
 gluons with $x'\gg x$ act as {\em frozen color sources emitting
   gluons at the scale $x$}. Because these sources may overlap in the
 transverse plane, their color charges add coherently, giving rise to
 a large color charge density. The {\em average} color charge density
 is zero by gauge symmetry but fluctuations in the color charge
 density are nonzero and increase rapidly with $1/x$.

 These considerations are at the heart of the CGC reformulation of
 BFKL evolution. However, in contrast to the original BFKL
 formulation, the CGC formalism \cite{MV,Balit1,Kovch3,CGC1,CGC2} 
 includes {\em non--linear} effects which appear when
 the gluon density becomes large.

 The quantity which controls gluon interactions in the IMF is their
 {\em occupation number}--the number of gluons of a given color per
 unit transverse phase--space and per unit rapidity,
 \begin{eqnarray}\label{phidef}
   n(Y,k_\perp,b_\perp)\,\equiv\,\frac{(2\pi)^3}{2(N_c^2-1)}\, \,\frac{{\rm
       d} N_g}{\rmd Y\, {\rm d}^2\k_\perp\,{\rm d}^2\b_\perp}\;,
\end{eqnarray}
where the impact parameter $\b_\perp$ is the gluon position in the
transverse plane. If $n\ll 1$, the system is dilute and gluon
interactions are negligible. When $n\sim\order{1}$ gluons start
overlapping, but their interactions are suppressed by $\abar\ll 1$.
The interaction strength becomes of order one when $n\sim
\order{1/\abar}$. It is then that non--linear effects become
important leading to gluon saturation.

Gluon occupancy is further amplified if instead of a proton we
consider a large nucleus with atomic number $A\gg 1$.  The transverse
area scales like $A^{2/3}$, and the gluon occupation number scales as
$A^{1/3}$. Thus, for a large nucleus, saturation effects become
important at larger values of $x$ than for a proton.

To be more specific, let us discuss a simplified version of the
non--linear evolution equation for the occupation number. Consider an
elementary increment in rapidity: $Y\to Y+\rmd Y$. Each preexisting
gluon in the hadron has a probability $\abar\rmd Y$ to emit an
additional soft gluon -- the average increase is $\rmd n\sim \abar
n\rmd Y$.  Further, the emission vertex is non-local in $k_\perp$
because the transverse momentum of the parent gluon is shared among
the two daughter gluons. At high-energies, this non-locality is well
approximated as a {\em diffusion} in the logarithmic variable $t\equiv
\ln(k_\perp^2)$. Finally, two preexisting gluons can merge and produce
a single final gluon with rapidity $Y+\rmd Y$. This process is
quadratic in $\abar n$ and leads to a negative term in the evolution
equation. Adding up these three effects, one obtains the evolution
equation
\begin{equation}
 \frac{\partial n}{\partial Y}
\simeq
\omega\abar n
 \,+\,\chi\abar\partial_t^2 n\,-\,\beta \abar^2 n^2\; ,
 \label{eq:GLR}
\end{equation}
where $\omega$, $\chi$, and $\beta$ are numbers of order unity. This
equation mimics the BFKL equation \cite{BFKL} if one drops the term 
quadratic in $\abar n$ in the r.h.s., and mimics its non-linear extensions
that include saturation if one keeps the quadratic term. 

\subsection{Generic features of gluon saturation}
Although a toy model, \eqnum{eq:GLR} captures essential features of
saturation\footnote{Our discussion oversimplifies the mechanism for
  gluon saturation. In the saturation regime, gluons form
  configurations that screen their color charge over transverse scales
  $\sim 1/Q_s$ \cite{IancuM1,screening}.  Thus, gluons with momentum
  $k_\perp\simle Q_s$ are emitted from a quasi-neutral patch of color
  sources, and their occupation number grows only {\em linearly} in
  $Y$ \cite{Muell3,IancuM1} -- much slower than the exponential growth
  in the region $k_\perp\gg Q_s$.}.  If  $\abar n\ll 1$, one can neglect the quadratic term, and eq.~(\ref{eq:GLR})
predicts an exponential growth in $Y$ of the gluon occupation number.
But when $\abar n\sim 1$, the negative non-linear term turns on and
tames the growth.  In fact, eq.~(\ref{eq:GLR}) has a fixed point at
$n=\omega/(\beta\abar)$ where its r.h.s. vanishes--the evolution stops
when this value of $n$ is reached, resulting in gluon saturation in
the spirit of early works~\cite{saturation}.

Eq.~(\ref{eq:GLR}) also reveals the emergence of a transverse momentum
scale $Q_s(Y)$ that characterizes saturation. This scale is the
$k_\perp$ where gluon occupancy becomes of $\order{1/\abar}$.  As a
function of transverse momentum, the occupation number $n(Y,k_\perp)$
is $\order{1/\abar}$ if $k_\perp \lesssim Q_s(Y)$, and decreases
rapidly above $Q_s(Y)$ ($n\propto 1/k_\perp^2$ for $k_\perp\gg
Q_s(Y)$). The shape of $n(Y,k_\perp)$ as a function of $k_\perp$ is
known as the {\sl saturation front}. Note that $Q_s$, the typical
gluon transverse momentum at the rapidity $Y$, increases with
energy and becomes a semi-hard scale ($Q_s(Y)\gg\Lam$) at sufficiently high energy.

Eq.~(\ref{eq:GLR}) belongs to the generic class of {\em
  reaction-diffusion processes}~\cite{Saarl1}. These are processes
where an entity can hop to neighboring locations (diffusion term
$\chi\abar\partial_t^2 n$), can split into two identical entities
(the term $\omega\abar n$), and where two entities can merge into a single
one (the term $\beta \abar^2 n^2$). In the limit of large occupation
numbers, these processes admit the mean field description of
eq.~(\ref{eq:GLR}), which in the context of statistical physics is
known as the {\em Fisher-Kolmogorov-Petrovsky-Piscounov (FKPP)
  equation}~\cite{Saarl1}.  In QCD, the closest equation of this type
is the Balitsky-Kovchegov (BK) equation~\cite{Kovch3}. The
correspondence between the BK and FKPP equations, originally noticed
in \cite{FKPP}, clarifies the properties of saturation fronts in QCD
in analogy with known properties of reaction--diffusion processes.

A crucial property is the emergence of {\em traveling waves}. The
saturation front generated by this equation propagates without
distortion at constant speed; one has $n(Y,t)=n(t-\lambda_s Y)$ with
$\lambda_s$ a constant. This property has been verified in numerical
studies of the BK equation~\cite{BKnum,RummuW1} and in analytic
studies of the BFKL equation in the presence of a saturation
boundary~\cite{BFKLboundary,Trian1}. It provides a natural explanation
of the {\em geometric scaling} phenomenon observed in the HERA
data~\cite{StastGK1,GelisPSS1} (see section \ref{sec:coll-DIS}).  In
QCD, a front moving with constant speed $\lambda_s$ is equivalent to
the saturation momentum increasing exponentially with $Y$,
 \begin{eqnarray}\label{Qsat}
   Q_s^2(Y)\simeq Q_0^2\,\rme^{\lambda_s Y}\quad{\rm with}\quad
   \lambda_s \approx 4.9\,\abar\; ,
\end{eqnarray}
where $Q_0$ is some non-perturbative initial scale. For a large
nucleus, $Q_0^2$ scales like $A^{1/3}$ as does $Q_s^2(Y)$ for any $Y$.
This form of the saturation momentum is modified to $Q_s^2=Q_0^2\,
e^{\sqrt{\lambda(Y+Y_0)}}$ when the running of the strong coupling is
taken into account; see ~\cite{Trian1} for a detailed study of higher
order effects on the energy dependence of $Q_s$.

\subsection{The Color Glass Condensate}
The CGC is an {\em effective field theory (EFT)} based on the
separation of the degrees of freedom into fast frozen color sources
and slow dynamical color fields~\cite{MV}. A {\em renormalization
  group equation} --the JIMWLK equation \cite{CGC1,CGC2}-- ensures the
independence of physical quantities with respect to the cutoff that
separates the two kinds of degrees of freedom.

The fast gluons with longitudinal momentum $k^+>\Lambda^+$ are frozen
by Lorentz time dilation in configurations specified by a color
current $J^\mu_a \equiv \delta^{\mu +}\rho^a$, where
$\rho^a(x^-,x_\perp)$ is the corresponding color charge density. On
the other hand, slow gluons with $k^+<\Lambda^+$ are described by the
usual gauge fields $A^\mu$ of QCD. Because of the hierarchy in $k^+$
between these two types of degrees of freedom, they are coupled
eikonaly by a term $J_\mu A^\mu$.  The fast gluons thus act as
sources for the fields that represent the slow gluons. Although it is
frozen for the duration of a given collision, the color source density
$\rho^a$ varies randomly event by event. The CGC provides a gauge
invariant distribution $W_{\Lambda^+}[\rho]$, which gives the
probability of a configuration $\rho$. This functional encodes all the
correlations of the color charge density at the cutoff scale
$\Lambda^+$ separating the fast and slow degrees of freedom. Given
this statistical distribution, the expectation value of an operator
at the scale $\Lambda^+$ is given by
\begin{equation}
\left<{\cal O}\right>_{\Lambda^+}\equiv
\int\big[D\rho\big]\;W_{\Lambda^+}\big[\rho\big]\;{\cal O}\big[\rho\big]\; ,
\end{equation}
where ${\cal O}[\rho]$ is the expectation value of the operator for a
particular configuration $\rho$ of the color sources.

The power counting of the CGC EFT is such that in the saturated regime
the sources $\rho$ are of order $g^{-1}$. Attaching an additional
source to a given Feynman graph does not alter its order in $g$; the
vertex where this new source attaches to the graph is compensated by
the $g^{-1}$ of the source. Thus, computing an observable at a certain
order in $g^2$ requires the resummation of all the contributions
obtained by adding extra sources to the relevant graphs.  The leading
order in $g^2$ is given by a sum of tree diagrams, which can be
expressed in terms of classical solutions of the Yang-Mills equations.
Moreover, for inclusive observables~\cite{classical}, these classical
fields obey a simple boundary condition: they vanish when $t\to
-\infty$.

Next-to-leading order (NLO) computations in the CGC EFT involve a sum
of one-loop diagrams embedded in the above classical field. To prevent
double counting, momenta in loops are required to be below the cutoff
$\Lambda^+$. This leads to a logarithmic dependence in $\Lambda^+$ of
these loop corrections. These logarithms are large if $\Lambda^+$
is well above the typical longitudinal momentum scale of the
observable considered, and must be resummed.

For gluon correlations inside the the hadron wavefunction and also for
sufficiently inclusive observables in a collision, the leading
logarithms are universal and can be absorbed into a redefinition of
the distribution $W_{\Lambda^+}[\rho]$ of the hard sources. The
evolution of $W_{\Lambda^+}[\rho]$ with $\Lambda^+$ is governed by the
functional JIMWLK equation
\begin{eqnarray}
 \frac{\del\, W_{\Lambda^+}[\rho] }{{\del \ln(\Lambda^+)}}=
-{\cal H}\left[\rho,\frac{\delta}{ {\delta \rho}}\right]\,W_{\Lambda^+}[\rho]\;,
\end{eqnarray}
where ${\cal H}$ is known as the JIMWLK Hamiltonian. This operator
contains up to two derivatives $\del/\del\rho$, and arbitrary powers
in $\rho$. Its explicit expression can be found in
refs.~\cite{CGC1,CGC2,reviews1}.  The derivation of the JIMWLK
equation will be sketched in the section \ref{sec:coll-DIS}.

Numerical studies of JIMWLK evolution were performed in
\cite{RummuW1,KovchKRW1}.  An analytic, albeit formal, solution to the
JIMWLK equation was constructed in \cite{BlaizIW1} in the form of a
path integral.  Alternatively, the evolution can can be expressed as
an infinite hierarchy of coupled non-linear equations for $n$-point
Wilson line correlators--often called the Balitsky
hierarchy~\cite{Balit1}.  In this framework, the BK equation is a mean
field approximation of the JIMWLK evolution, valid in the limit of a
large number of colors $N_c\to\infty$. Numerical studies of the JIMWLK
equation~\cite{RummuW1,KovchKRW1} have found only small differences
with the BK equation.

Let us finally comment on the initial condition for the JIMWLK
equation which is also important in understanding its derivation. The
evolution should start at some cutoff value in the longitudinal
momentum scale $\Lambda^+_0$ at which the saturation scale is already
a (semi)hard scale, say $Q_{s0}\gtrsim 1$~GeV, for perturbation theory
to be applicable.  The gluon distribution at the starting scale is in
general non--perturbative and requires a model.  A physically
motivated model for the gluon distribution in a large nucleus is the
McLerran-Venugopalan model~\cite{MV}. In a large nucleus, there is a
window in rapidity where evolution effects are not large but $x$ is
still sufficiently small for a probe not to resolve the longitudinal
extent of the nucleus. In this case, the probe ``sees'' a large number
of color charges, proportional to $A^{1/3}$.  These charges add up to
form a higher dimensional representation of the gauge group, and can
therefore be treated as classical color
distributions~\cite{MV,JeonV1}.  Further, the color charge
distribution $W_{\Lambda_0^+}[\rho]$ is a Gaussian
distribution\footnote{There is a additional term, corresponding to the
  cubic Casimir; which is parametrically suppressed for large
  nuclei~\cite{JeonV2}. This term generates Odderon excitations in the
  JIMWLK/BK evolution~\cite{odderon}.} in $\rho$. The variance of this
distribution --the color charge squared per unit area-- is
proportional to $A^{1/3}$ and provides a semi-hard scale that makes
weak coupling computations feasible. In addition to its role in
motivating the EFT and serving as the initial condition in JIMWLK
evolution, the MV model allows for direct phenomenological studies in
p+A and A+A collisions in regimes where the values of $x$ are not so
small as to require evolution.

The dynamics of small x gluons in QCD may be universal in more than
one sense~\cite{McLer2}.  A weak form of this universality is that
their dynamics in both hadrons and nuclei is controlled only by the
saturation scale with its particular dependence on energy and nuclear
size. A stronger form of the universality is noticed in particular for
the solution of the BK equation with running coupling effects; the
saturation scale, for both hadrons and nuclei, at fixed impact
parameter, becomes the same asymptotically with increasing
energy~\cite{Muell7}.  The strongest form of the universality is that
the RG flows in the saturation regime have a fixed point corresponding
to universal ``critical'' exponents describing the behavior of
multi-parton correlation functions. As discussed further below, the RG
equations for high energy QCD lie in a wide class of
reaction-diffusion processes which have universal properties
remarkably close to those of spin glasses~\cite{FKPP1}.

\subsection{Advanced Theory topics}
Reaction--diffusion processes exhibit an extreme sensitivity to
{\em particle number fluctuations} \cite{sFKPP,MuellS2,IancuMM1},
generated by gluon splittings, which produce correlations among pairs
of gluons \cite{ploop}. This effect is of higher order in $\abar$, and
is linear in $n$ since it results from the splitting of a single
gluon.  Conversely, producing two gluons without this splitting leads
to a term that has one less power of $\abar$, but of order $n^2$.
Thus, the splitting contribution is important in the {\em dilute}
regime where $n\lesssim \abar$. Since, as mentioned previously, the
dynamics of the saturation front is driven by the BFKL growth of its
dilute tail, these fluctuations play an important role in the 2--gluon
density $\langle nn\rangle$ at high energy.

As manifest in \eqnum{eq:GLR}, this 2-gluon density enters
in the non--linear term leading to saturation of the single gluon
density.  Thus, gluon number fluctuations in the dilute regime can
strongly influence the approach towards saturation. For instance, 
as argued in \cite{sFKPP} for generic reaction--diffusion processes, and
independently in the QCD context\cite{MuellS2}, these
fluctuations reduce the (average) speed of the saturation front.
Besides making the value of $Q_s$ a fluctuating quantity, they tend to
wash out the geometric scaling property of the individual fronts
\cite{IancuMM1,ploop}.  Both effects are quantitatively
important, as shown by explicit numerical simulations within various
reaction--diffusion models (including those inspired by the QCD
dynamics at high energy \cite{IancuAST1}).

There is presently no general theory that includes BFKL ladders,
saturation and fluctuations. (In terms of Feynman graphs, this 
corresponds to resumming ``Pomeron loops'' diagrams to all orders
\cite{ploop}.) Attempts to construct such a theory
\cite{ploop,MuellSW3,ploop1} have led to incomplete formalisms that
are difficult to exploit in phenomenological applications.
Fortunately, the effect of these fluctuations is considerably reduced
by the running of the strong coupling \cite{DumitIPST1}, which tends
to postpone their importance to unrealistically large rapidities.
Thus, for practical applications at least up to LHC energies, a
sufficient theory for the approach towards saturation is the
leading--order mean-field evolution extended with running coupling
corrections. This theory has developed significantly as the
running--coupling version of the BK equation has been constructed in
\cite{BKrunning} and successfully applied to studies of the
phenomenology at HERA \cite{AlbacAMS1}.

\section{Collisions in the CGC framework}
The CGC is an effective theory for the wavefunction of a high-energy
hadron or nucleus.  In this section, we apply it, with particular
emphasis on factorization, to deeply inelastic scattering and hadronic
collisions. In A+A collisions, the formation of the Glasma and its key
features are emphasized.

\subsection{The CGC and DIS at small $x$}
\label{sec:coll-DIS}
\begin{figure*}[htb!]
 \begin{center}
 \includegraphics[width=0.25\textwidth]{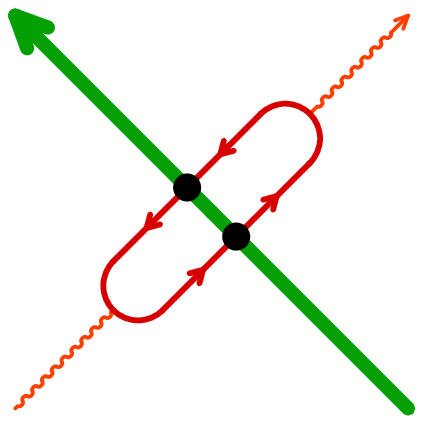}
 \hskip 20mm
 \includegraphics[width=0.25\textwidth]{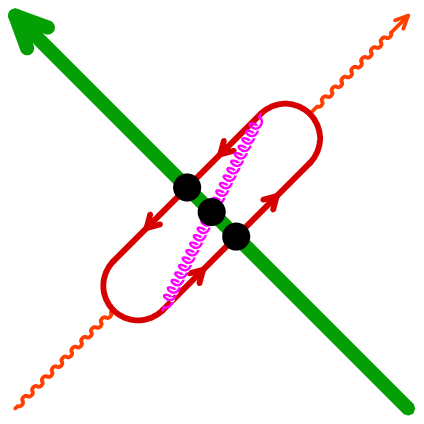}
\end{center}
\caption{\baselineskip 15pt  \label{fig:DIS-LO}\sl Left: leading Order (LO) contribution
  to DIS off the CGC. Right: NLO contribution.}
\end{figure*}

\begin{figure*}[htb!]
 \begin{center}
 \centerline{\epsfig{file=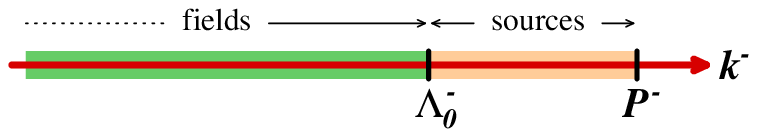,width=7cm}}
\vskip 2mm
 \centerline{\epsfig{file=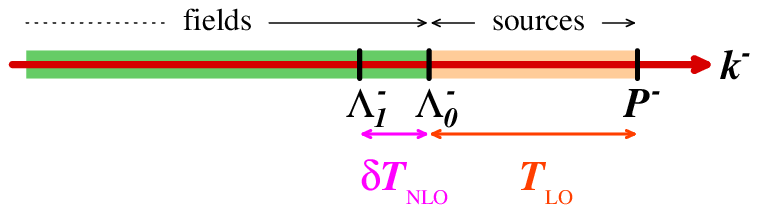,width=7cm}}
 \caption{\baselineskip 15pt  \label{fig:CGC-1}\sl Top: sources and fields in the CGC
   effective theory. Bottom: NLO correction from a layer of field
   modes just below the cutoff.}
\end{center}
\end{figure*}

At small $x$, corresponding to large Ioffe times~\cite{ioffe}, DIS is
characterized by the fluctuation of the virtual photon into a
quark--antiquark pair which then scatters off the hadronic or nuclear
target. The inclusive DIS cross-section can be expressed
as~\cite{NikolZ1}
\begin{equation}
\sigma_{\gamma^*T}
=\int_0^1 \rmd z\int \rmd^2\r_\perp
{
\left|\psi(z,\r_\perp)\right|^2
}\;
\sigma_{\rm dipole}(x,\r_\perp)
\; ,
\label{eq:X-DIS}
\end{equation}
where $\psi(z,\r_\perp)$ is the $q\bar{q}$ component of the
wave-function of the virtual photon (known from QED) and $\sigma_{\rm
  dipole}(x,\r_\perp)$ is the QCD ``dipole'' cross-section for the
quark-antiquark pair to scatter off the target.  This process is shown
in figure~\ref{fig:DIS-LO}, where we have assumed that the target
moves in the $-z$ direction. In the leading order (LO) CGC description
of DIS, the target is described, as illustrated in
figure~\ref{fig:CGC-1}, as static sources with $k^->\Lambda_0^-$. The
field modes do not contribute at this order.

Employing the optical theorem, $\sigma_{\rm dipole}(x,\r_\perp)$ can
be expressed in terms of the forward scattering amplitude ${\bs
  T(\x_\perp,\y_\perp)}$ of the $q\bar{q}$ pair at LO as
\begin{equation}
\sigma_{\rm dipole}^{_{\rm LO}}(x,\r_\perp) 
= 
2 \int \; d^2{\bf b}\;\int \; [D\rho] W_{\Lambda_0^-}[\rho]\;
 {\bs T}_{_{\rm LO}}(\b+\frac{\r_\perp}{2}, \b - \frac{\r_\perp}{2})\; ,
\label{eq:opt-thm-LO}
\end{equation}
where, for a fixed configuration of the target color sources
\cite{McLerV4,Venug1}
\begin{equation}
{ {\bs T}_{_{\rm LO}}(\x_\perp,\y_\perp)}
=
1-\frac{1}{N_c}\,{\rm tr}\,( U(\x_\perp)U^\dagger(\y_\perp))\; ,
\label{eq:Wilson-amp}
\end{equation}
with $U(\x_\perp)$ a Wilson line representing the interaction between
a quark and the color fields of the target, defined to be
\begin{equation}
U(\x_\perp)
=
 {\rm T}\,\exp i{g}\int^{1/x P^-} 
 dz^+\,{{\cal A}^-(z^+,\x_\perp)}\; .
\end{equation}
In this formula, ${\cal A}^-$ is the minus component of the gauge
field generated (in Lorenz gauge) by the sources of the target; it is
obtained by solving classical Yang-Mills equations with these sources.
The upper bound $xP^-$ (where $P^-$ is the target longitudinal
momentum and $x$ the kinematic variable defined in
figure~\ref{fig:DISkinematics}) indicates that source modes with
$k^-<xP^-$ do not contribute to this scattering amplitude. Thus if the
cutoff $\Lambda_0^-$ of the CGC EFT is lower than $xP^-$, ${\bs
  T}_{_{\rm LO}}$ is independent of $\Lambda_0^-$.

However, when $\Lambda_0^-$ is larger than $xP^-$, the dipole
cross-section is in fact independent of $x$ (since the CGC EFT does
not have source modes near the upper bound $xP^-$) and depends on the
unphysical parameter $\Lambda_0^-$. As we shall see now, this is
related to the fact that eq.~(\ref{eq:opt-thm-LO}) is incomplete and
receives large corrections from higher order diagrams. Consider now
the NLO contributions (one of them is shown in the right panel in
figure \ref{fig:DIS-LO}) with gauge field modes in the slice
$\Lambda_1^-\le k^- \le \Lambda_0^-$ (see figure~\ref{fig:CGC-1}).
An explicit computation of the contribution of field modes in this
slice gives
\begin{equation}
    {\delta {\bs T}_{_{\rm NLO}}(\x_\perp,\y_\perp)}
    =
    \ln\left(\frac{\Lambda_0^-}{\Lambda_1^-}\right)\;
    {{\cal H}}\;
    {{\bs T}_{_{\rm LO}}(\x_\perp,\y_\perp)}\; ,
    \label{eq:DIS-NLO}
\end{equation}
where ${\cal H}$ is the JIMWLK Hamiltonian. All dependence on the
cutoff scales is in the logarithmic prefactor alone. This Hamiltonian
has two derivatives with respect to the classical field ${\cal
  A}\sim{\cal O}(1/g)$; ${\cal H}\,{\bs T}_{_{\rm LO}}$ is of order
$\alpha_s {\bs T}_{_{\rm LO}}$ and therefore clearly an NLO
contribution. However, if the new scale $\Lambda_1^-$ is such that
$\alpha_s\ln(\Lambda_0^-/\Lambda_1^-)\sim 1$, this NLO term becomes
comparable in magnitude to the LO contribution. Averaging the sum of
the LO and NLO contributions over the distribution of sources at the
scale $\Lambda_0^-$, one obtains
\begin{eqnarray}
    \int[D\rho]\;W_{\Lambda_0^-}[\rho]\;
    \left({\bs T}_{_{\rm LO}}+\delta{\bs T}_{_{\rm NLO}}\right)=
    \int[D\rho]\;W_{\Lambda_1^-}[\rho]\;{\bs T}_{_{\rm LO}}\; ,
\label{eq:int-by-parts}
\end{eqnarray}
where $W_{\Lambda_1^-}\equiv (1+\ln(\Lambda_0^-/\Lambda_1^-)\,{\cal
  H})\,W_{\Lambda_0^-}$. We have shown here that the NLO correction
from quantum modes in the slice $\Lambda_1^-\le k^- \le \Lambda_0^-$
can be absorbed in the LO term, provided we now use a CGC effective
theory at $\Lambda_1^-$ with the modified distribution of sources
shown in eq.~(\ref{eq:int-by-parts}).  In differential form, the
evolution equation of the source distribution,
\begin{equation}
\frac{\partial}{\partial\ln(\Lambda^-)}W_{\Lambda^-}
=-{\cal H}\,W_{\Lambda^-}\; ,
\label{eq:JIMWLK-diff}
\end{equation}
is the JIMWLK equation.

Repeating this elementary step, one progressively resums quantum
fluctuations down to the scale $k^-\sim xP^-$. Thanks to
eq.~(\ref{eq:int-by-parts}), the result of this resummation for the
dipole cross-section is formally identical to
eq.~(\ref{eq:opt-thm-LO}), except that the source distribution is
$W_{xP^-}$ instead of $W_{\Lambda_0^-}$. Note that if one further
lowers the cutoff below $xP^-$, the dipole cross-section remains
unchanged.

\subsection{The CGC in p+A collisions}
\begin{figure}[htb!]
  \begin{center}
    \includegraphics[width=0.36\textwidth]{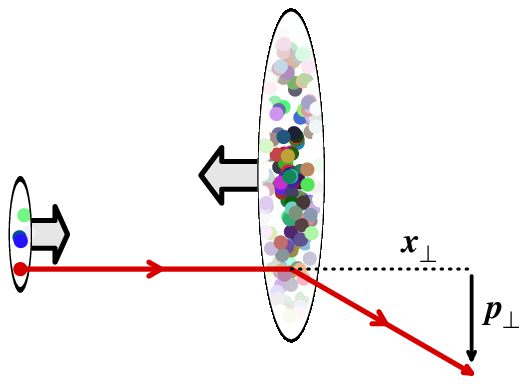}
    \hskip 5mm
    \includegraphics[width=0.54\textwidth]{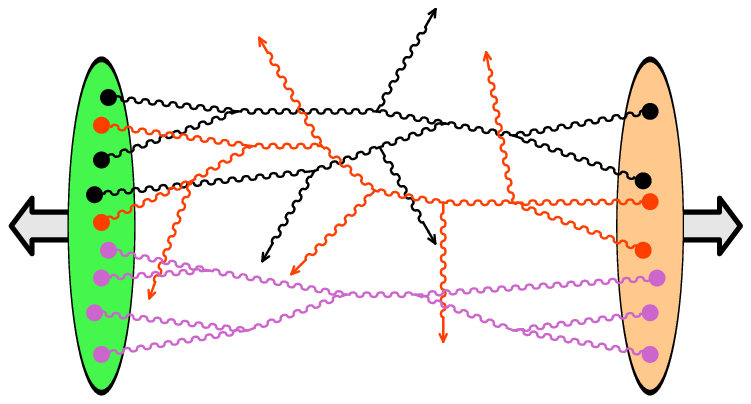}
  \end{center}
  \caption{\baselineskip 15pt  \label{fig:pA}\sl Left: sketch of a proton-nucleus
    collision. Right: example of leading order contribution in a
    nucleus-nucleus collision.}
\end{figure}
Collisions between a dilute hadron projectile and a dense hadron
target can be studied semi-analytically in the CGC framework. The
archetype of such collisions is a proton-nucleus collision. However,
the dilute-dense treatment also applies to proton-proton collisions
for measurements at forward rapidities where the wavefunction of one
of the projectiles is probed at large $x$ and that of the other at
small $x$.  These asymmetrical collisions can be treated using
conventional parton distributions for the proton and the CGC for the
nucleus.

The simplest quantity to compute in this context is the single
particle inclusive spectrum. In the CGC framework, this process is due
to the scattering of a parton from the proton off the color field of
the target nucleus, as illustrated in figure~\ref{fig:pA} (left). When
the parton is a quark, the amplitude for the scattering is
proportional to the Fourier transform of a Wilson line,
\begin{equation}
{\cal M}_{_{\rm LO}}\propto \int d^2\x_\perp\;
e^{i\p_\perp\cdot\x_\perp}\; U(\x_\perp)\; .
\label{eq:pA-1incl-q}
\end{equation}
Squaring this amplitude, summing over the color of the quark in the
final state and averaging over the color of the incoming quark, we get
\begin{equation}
\left|{\cal M}\right|^2_{_{\rm LO}}
\propto
      \int d^2\b \; d^2\r_\perp\;
      e^{i\p_\perp\cdot\r_\perp}\;
      {\bs T}_{_{\rm LO}}\left(\b+\frac{\r_\perp}{2},\b-\frac{\r_\perp}{2}\right)\; ,
\label{eq:tmp1}
\end{equation}
where ${\bs T}_{_{\rm LO}}$ is the dipole scattering amplitude (see
eq.~(\ref{eq:Wilson-amp})) that already appeared in our discussion of
DIS. Because the same quantity appears here, the treatment of the NLO
corrections we described in the DIS case is similarly applicable; one
integrates out softer modes by lowering the cutoff of the CGC EFT
letting $W_{\Lambda^-}[\rho]$ evolve according to the JIMWLK equation.
The scale to which one evolves the cutoff is $\Lambda^-=xP^-$ with
$x=(p_\perp/\sqrt{s})\exp(-y)$ where $p_\perp$ is the transverse
momentum of the scattered parton and $y$ its rapidity. Therefore, in
the CGC framework, the cross-section for this process is simply the
Fourier transform of the dipole cross-section $\sigma_{\rm
  dipole}(x,\r_\perp)$. In p+A collisions, final states containing a
photon or a lepton pair can be similarly expressed in terms of the
same Fourier transform \cite{photon}.

An additional contribution to the single inclusive particle spectrum
in a p+A collision is due to an incoming gluon instead of a quark. The
treatment is nearly identical to the incoming quark case, except that
in eq.~(\ref{eq:pA-1incl-q}) one must replace the Wilson line in the
fundamental representation by a Wilson line in the adjoint
representation. 

Similar calculations can be performed for processes with more
complicated final states, such as the production of a quark-antiquark pair.
Although this observable has a more complicated expression (containing terms that are the product of four Wilson lines) its NLO
corrections still comply with eq.~(\ref{eq:DIS-NLO}), which ensures
their factorization into the distribution of sources
$W_{\Lambda^-}[\rho]$. The crucial ingredient for factorization to
work is to consider an observable that is sufficiently inclusive to
allow the corresponding final state to be accompanied by an arbitrary
number of gluons. Any restriction on associated gluon radiation will
not permit NLO corrections to factorize simply in the distribution of
sources.

A much more limited but widely used form of factorization is $k_\perp$
factorization~\cite{KTfact} in terms of $k_\perp$ dependent
unintegrated quark and gluon distributions of the projectile and
target. Within the CGC framework, these results can be reproduced for
gluon~\cite{KovchM3,DumitM1} and heavy quark~\cite{GelisV1}
distributions at large transverse momenta $k_\perp \geq Q_s$; however,
at smaller transverse momenta, $k_\perp$ factorization is broken even
at leading order~\cite{KrasnV2,BlaizGV2,NikolSZ1}.

\subsection{Shattering CGCs in A+A collisions}
\label{sec:AA}
Collisions between two nuclei (``dense-dense'' scattering) are
complicated to handle on the surface. However, in the CGC framework,
because the wave functions of the two nuclei are saturated, the
collision can be treated as the collision of classical fields. This insight significantly simplifies the treatment of A+A scattering. 
The classical fields are coupled to fast partons of each nucleus respectively described by the external
current $J^\mu=\delta^{\mu+}\rho_1+\delta^{\mu-}\rho_2$.  The source
densities of fast partons $\rho_{1,2}$ are both parametrically of
order $1/g$, which implies that graphs involving multiple sources from
both projectiles must be resummed. (See the right panel of
figure~\ref{fig:pA} for an illustration.)

At leading order, inclusive observables\footnote{Exclusive observables
  may also be expressed in terms of solutions of the same Yang-Mills
  equations, but with more complicated boundary conditions than for
  inclusive observables.} depends on the retarded classical color
field ${\cal A}^\mu$, which solves the Yang-Mills equations $[{\cal
  D}_\mu,{\cal F}^{\mu\nu}]=J^\nu$ with the boundary condition
$\lim_{x^0\to -\infty}{\cal A}^\mu =0$. Among the observables to which
this result applies is the expectation value of the energy-momentum
tensor at early times after the collision. At leading order,
  \begin{equation}
T^{\mu\nu}_{_{\rm LO}}
=
\frac{1}{4}g^{\mu\nu}\,{{\cal F}^{\lambda\sigma}{\cal F}_{\lambda\sigma}}
-{{\cal F}^{\mu\lambda}{\cal F}^\nu{}_\lambda}\; ,
\end{equation}
where ${\cal F}^{\mu\nu}$ is the field strength of the classical field
${\cal A}^\mu$.

Although A+A collisions are more complicated than e+A or p+A
collisions, one can still factorize the leading higher order
corrections into the evolved distributions $W_{\Lambda^-}[\rho_1]$ and
$W_{\Lambda^+}[\rho_2]$. At the heart of this factorization is a
generalization of eq.~(\ref{eq:DIS-NLO}) to the case where the two
projectiles are described in the CGC framework \cite{factorization}.
When one integrates out the field modes in the slices
$\Lambda_1^\pm\le k^\pm\le \Lambda_0^\pm$, the leading correction to
the energy momentum tensor is
\begin{equation}
{\delta T^{\mu\nu}_{_{\rm NLO}}}
    =
    \Big[
    \ln\left(\frac{\Lambda_0^-}{\Lambda_1^-}\right)\,{{\cal H}_1}
    +
    \ln\left(\frac{\Lambda_0^+}{\Lambda_1^+}\right)\,{{\cal H}_2}
    \Big]\;{T^{\mu\nu}_{_{\rm LO}}}\; ,
\label{eq:AA-NLO}
\end{equation}
where ${\cal H}_{1,2}$ are the JIMWLK Hamiltonians of the two nuclei
respectively. What is crucial here is the absence of mixing between
the coefficients ${\cal H}_{1,2}$ of the logarithms of the two
projectiles; they depend only on $\rho_{1,2}$ respectively and not on
the sources of the other projectile.  Although the proof of this
expression is somewhat involved, the absence of mixing is deeply
rooted in causality.  The central point is that because the duration
of the collision (which scales as the inverse of the energy) is so
brief, soft radiation must occur before the two nuclei are in causal
contact. Thus logarithms associated with this radiation must have
coefficients that do not mix the sources of the two projectiles.

Following the same procedure for eq.~(\ref{eq:AA-NLO}), as for the e+A
and p+A cases, one obtains for the energy-momentum tensor in an A+A
collision the expression
\begin{equation}
\left<T^{\mu\nu}\right>_{_{\rm LLog}}
=
\int 
\big[D{\rho_{_1}}\,D{\rho_{_2}}\big]
\;
{ W_1\,[\rho_{_1}\big]}\;
{ W_2\big[\rho_{_2}\big]}
\;
T^{\mu\nu}_{_{\rm LO}}\; .
\label{eq:Tmunu}
\end{equation}
This result can be generalized to multi-point correlations of the
energy-mo\-men\-tum tensor,
\begin{eqnarray}
\left<T^{\mu_1\nu_1}(x_1)\cdots T^{\mu_n\nu_n}(x_n)\right>_{_{\rm LLog}}
&=&
\int 
\big[D{\rho_{_1}}\,D{\rho_{_2}}\big]
\;
{ W_1\,[\rho_{_1}\big]}\;
{ W_2\big[\rho_{_2}\big]}
\nonumber\\
&&\qquad\times
T^{\mu_1\nu_1}_{_{\rm LO}}(x_1)\cdots T^{\mu_n\nu_n}_{_{\rm LO}}(x_n)\; .
\label{eq:Tmunu2}
\end{eqnarray}
In this expression, all the correlations between the energy-momentum
tensor at different points are from the distributions
$W_{1,2}[\rho_{1,2}]$. Thus, the leading correlations are already
built into the wavefunctions of the projectiles prior to the
collision.

\begin{figure}[htb!]
\begin{center}
\resizebox*{5.8cm}{!}{\includegraphics{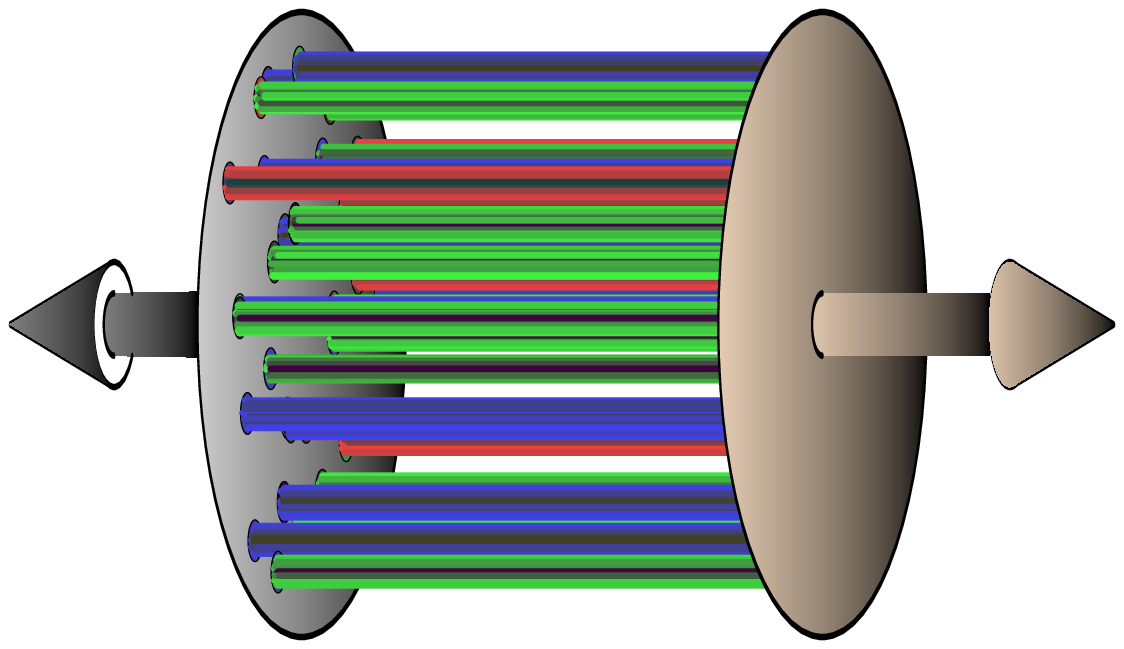}}
\hskip 1mm
\resizebox*{5.9cm}{!}{\includegraphics{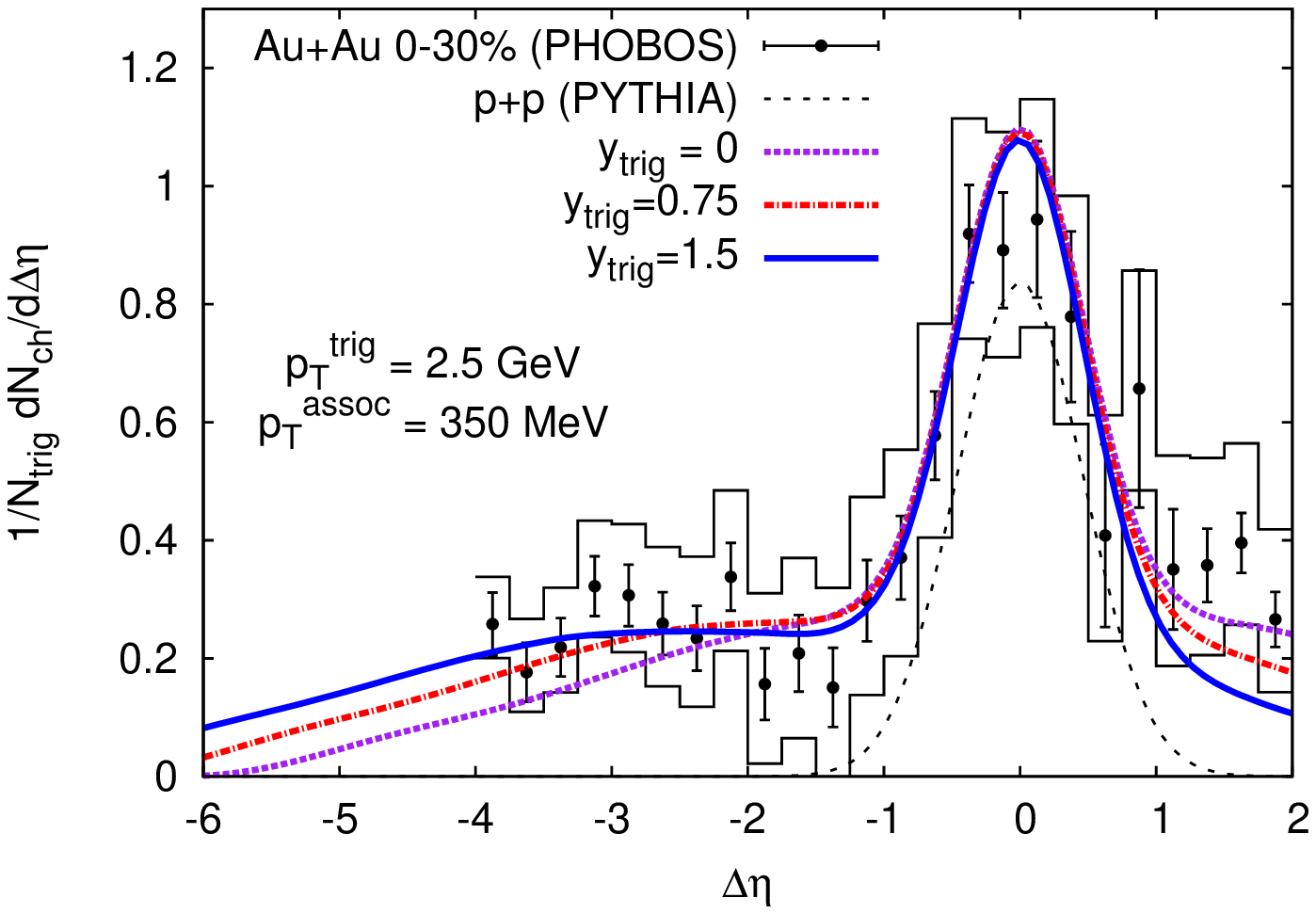}}
\end{center}
\caption{\baselineskip 15pt \label{fig:flux-tubes}\sl Left: Gauge
  field configurations in the form of ``flux tubes'' of longitudinal
  chromo-electric and chromo-magnetic fields screened on transverse
  scales $1/Q_s$. Right: Model comparison~\cite{DusliGLV1} to long
  range rapidity correlations measured by the PHOBOS
  collaboration~\cite{Alvera1}.}
\end{figure}

The expressions in eqs.~(\ref{eq:Tmunu}) and (\ref{eq:Tmunu2}) are
valid for proper times $\tau\sim 1/Q_s$ after the heavy ion collision.
The energy-momentum tensor, for each configuration of sources
$\rho_{1,2}$ is determined by solving classical Yang--Mills equations
to compute the gauge fields ${\cal A}_\mu^{\rm cl.}[\rho_1,\rho_2]$ in
the forward light cone with initial conditions determined by the
classical CGC fields of each of the nuclei at
$\tau=0$~\cite{class0,KrasnV2,KrasnNV1,Lappi,class1}.
The corresponding non-equilibrium matter, with high occupation numbers
$\sim 1/\alpha_s$ is called the Glasma~\cite{LappiM1}. The Glasma
fields at early times are longitudinal chromo-electric and
chromo-magnetic fields that are screened at distances $~1/Q_s$ in the
transverse plane of the collision. As a consequence, the matter
produced can be visualized (see figure~\ref{fig:flux-tubes})
as comprising $R_A^2 Q_s^2$ color flux tubes of size $1/Q_s$, each
producing $1/\alpha_s$ particles per unit rapidity.  The flux tube
picture is supported by non-perturbative numerical solutions of the
classical Yang-Mills equations~\cite{LappiSV1}. The ``Glasma flux
tubes'' generate $n$-particle long range rapidity
correlations~\cite{DumitGMV1,factorization,DusliFV1,DusliGLV1}. These
distributions are negative binomial distributions~\cite{GelisLM1}.
They also carry topological charge~\cite{KharzKV1}; the resulting
dynamical topological ``sphaleron'' transitions may result in
observable metastable CP-violating domains~\cite{KharzMW1}.
 
The evolution of the Glasma into a thermalized Quark Gluon Plasma
(QGP) is not understood. An important ingredient is the role of
instabilities \cite{instabilities}. At early times, these arise at NLO
from terms that break the boost invariance of the LO
term~\cite{GelisLV2,factorization}.  The modification to the evolution
of the Glasma is obtained by solving 3+1-D Yang-Mills
equations~\cite{insta1} for the (now) rapidity dependent gauge fields
convolved with a distribution giving the spectrum of
fluctuations~\cite{FukusGM1}. While these effects may isotropize the
system, early thermalization may also require collisions whose role
still needs to be clarified~\cite{collisions}.

\section{Phenomenological applications of the CGC}

In this section, we will discuss the applications of the theoretical
formalism outlined in the previous section to analyze and {\sl
  predict} a wide range of phenomena ranging from DIS in e+p and e+A
collisions to the scattering of hadronic projectiles ranging from p+p
to p+A to A+A collisions. A unifying ingredient in many of the
applications is the dipole cross-section defined in
eq.~(\ref{eq:opt-thm-LO}), albeit, as apparent in the treatment of A+A
collisions, the fundamental ingredient is really the density matrix
$W_{Y}[\rho]$.  Because the JIMWLK equation
(eq.~(\ref{eq:JIMWLK-diff})) for this quantity is time consuming to
solve\footnote{Other unknowns include higher order corrections,
  initial conditions at low energy and impact--parameter dependence of
  distributions.}, many of the applications are in the context of
models of the dipole cross-section which incorporate key features of
saturation. These models provide an economical description of a wide
range of data with only a few parameters. A good compromise between
the full JIMWLK dynamics and models of the dipole cross-section is the
BK equation, which is a large $N_c$ realization of JIMWLK dynamics.
With the recent availability of the NLO BK equation, global analyzes
of data are in order. Much of our discussion below is in the context
of dipole models; improvements {\it a la} BK are highlighted wherever
available.

\subsection{DIS in e+p and e+A collisions}

\begin{figure}[htb!]
\begin{center}
  \resizebox*{!}{4.5cm}{\includegraphics{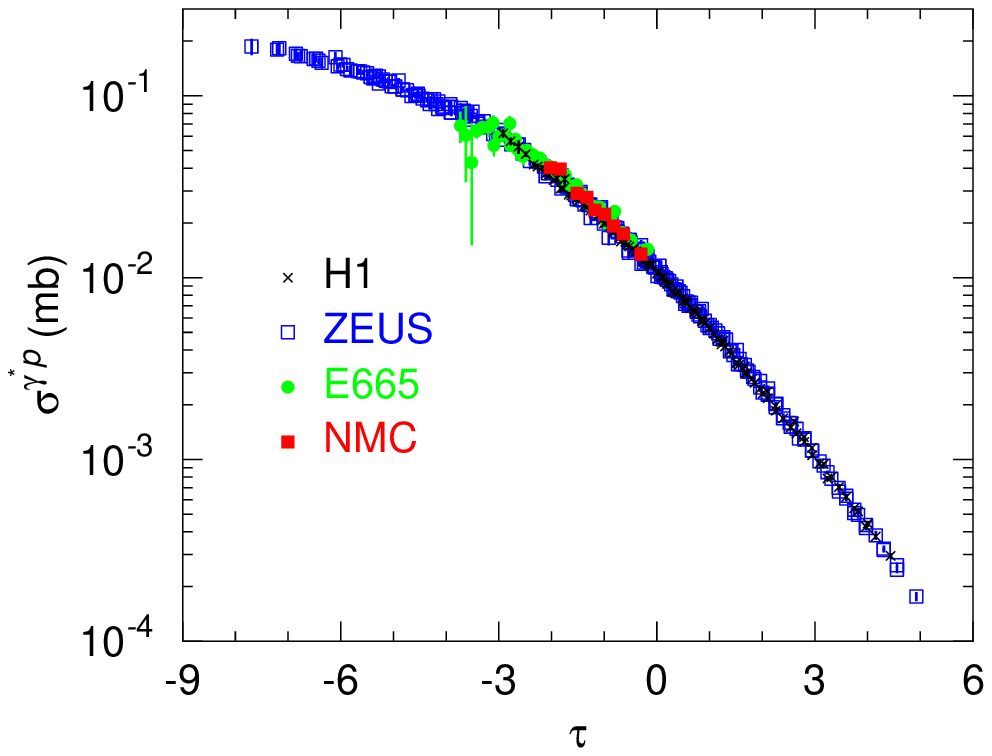}}
  \hskip 3mm
  \resizebox*{!}{4.5cm}{\includegraphics{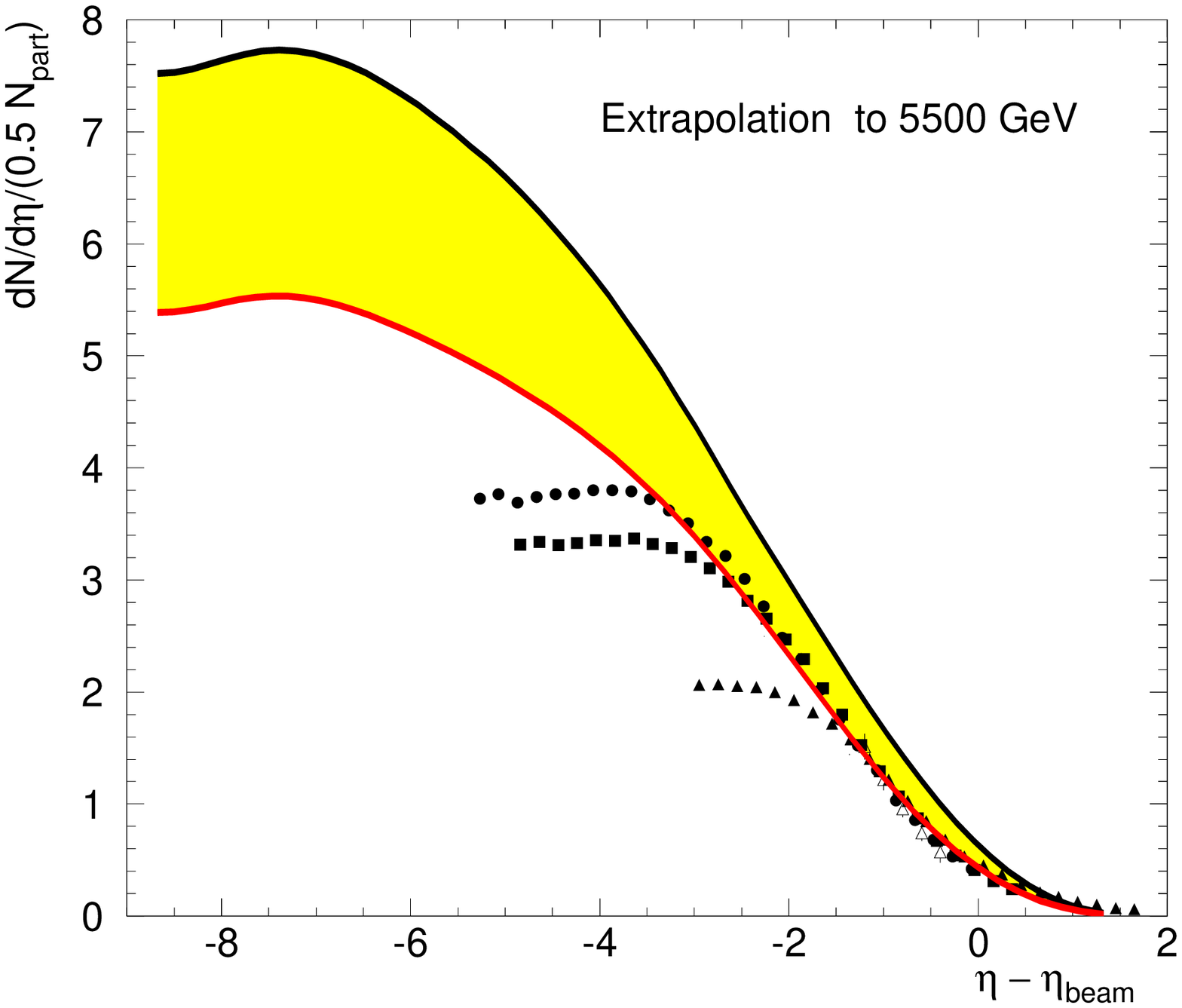}}
\end{center}
\caption{\baselineskip 15pt  \label{fig:geom}\sl Geometrical scaling and limiting
  fragmentation. Left: $\sigma_{\gamma^*p}$ data at HERA for $x\le
  0.01$ and all $Q^2$ up to 450~GeV$^2$; $\tau$ is the scaling
  variable, $\tau\equiv Q^2/Q^2_s(x)$ \cite{StastGK1,GelisPSS1}.
  Right: particle multiplicities for several collision energies at
  RHIC \cite{Backa1}, compared to the computation of \cite{GelisSV1}.}
\end{figure}

A remarkable observation~\cite{StastGK1} is that HERA
data~\cite{StastGK1,GelisPSS1} on the inclusive virtual photon-proton
cross section for $x\leq 0.01$ scale as a function of the ratio
$Q^2/Q_s^2(x)$; see the left part of figure \ref{fig:geom}. This
scaling is violated for larger values of $x$. Here $Q_s^2 = Q_0^2
(x_0/x)^\lambda$ is the saturation scale with $Q_0^2 = 1$ GeV$^2$,
$x_0 = 3\cdot 10^{-4}$ and $\lambda \approx 0.3$. This scaling is referred
to as ``geometrical scaling'', because the survival probability of the
color dipole that the virtual photon fluctuates into is close to unity
or zero respectively depending on whether the ratio of the saturation
radius ($\sim 1/Q_s$) to the size of the dipole (of size $\sim 1/Q$)
is large or small. Recall that the saturation radius denotes the
typical size of regions with strong color fields.
Geometric scaling has also been observed in inclusive diffraction,
exclusive vector meson production and deeply virtual Compton
scattering data at HERA~\cite{MarquS1}. In detail, the data also show
{\em violations} of geometric scaling, which can be interpreted as
consequences of BFKL diffusion \cite{IancuIM3}, non--zero quark masses
\cite{Soyez1} and possibly DGLAP evolution as well~\cite{KowalMW1}.
Note that the best scaling is obtained with a saturation scale that
behaves like $Q_s^2(x)\propto x^{-0.3}$, a slower $x$-dependence than
predicted by the LO BK equation. This discrepancy is resolved by a
resummed NLO computation of the saturation exponent \cite{Trian1}
which indeed gives $0.3$.

While geometrical scaling is very suggestive of the presence of
semi-hard dynamical scales in the proton, it is not conclusive in and
of itself~\cite{CaolaF1}; more detailed comparisons to the data are
essential.  Despite their simplicity, saturation
models~\cite{GBW,BarteGK1,IancuIM3,KowalT1,KowalMW1,Soyez1,forshaw,AlbacAMS1}
provide remarkably good descriptions of HERA data at small $x\le
0.01$. The free parameters are fixed from fits to the total
cross-section data alone; once these are fixed, the models {\sl
  predict} a large variety of results, including the longitudinal
($F_{_L}$), diffractive ($F_2^{_D}$), and charm ($F_2^c$) structure
functions, the virtual photon production of vector mesons
($\rho,\,J/\psi$), and the deeply virtual Compton scattering (DVCS).
The most recent analysis of inclusive data~\cite{AlbacAMS1} of DIS in
e+p collisions is quite sophisticated; the energy dependence is given
by the running--coupling BK equation, and the free parameters refer
solely to the initial conditions and to the proton transverse area.

The phenomenon of {\em hard diffraction} in DIS is particularly
sensitive to saturation. The simplest diffractive processes are events
in which the proton remains intact and a large gap in rapidity with no
particles extends between the rapidity of the proton and that of the
fragmentation products of the virtual photon.
\begin{figure*}[htb!]
  \begin{center}
    \includegraphics[width=0.45\textwidth]{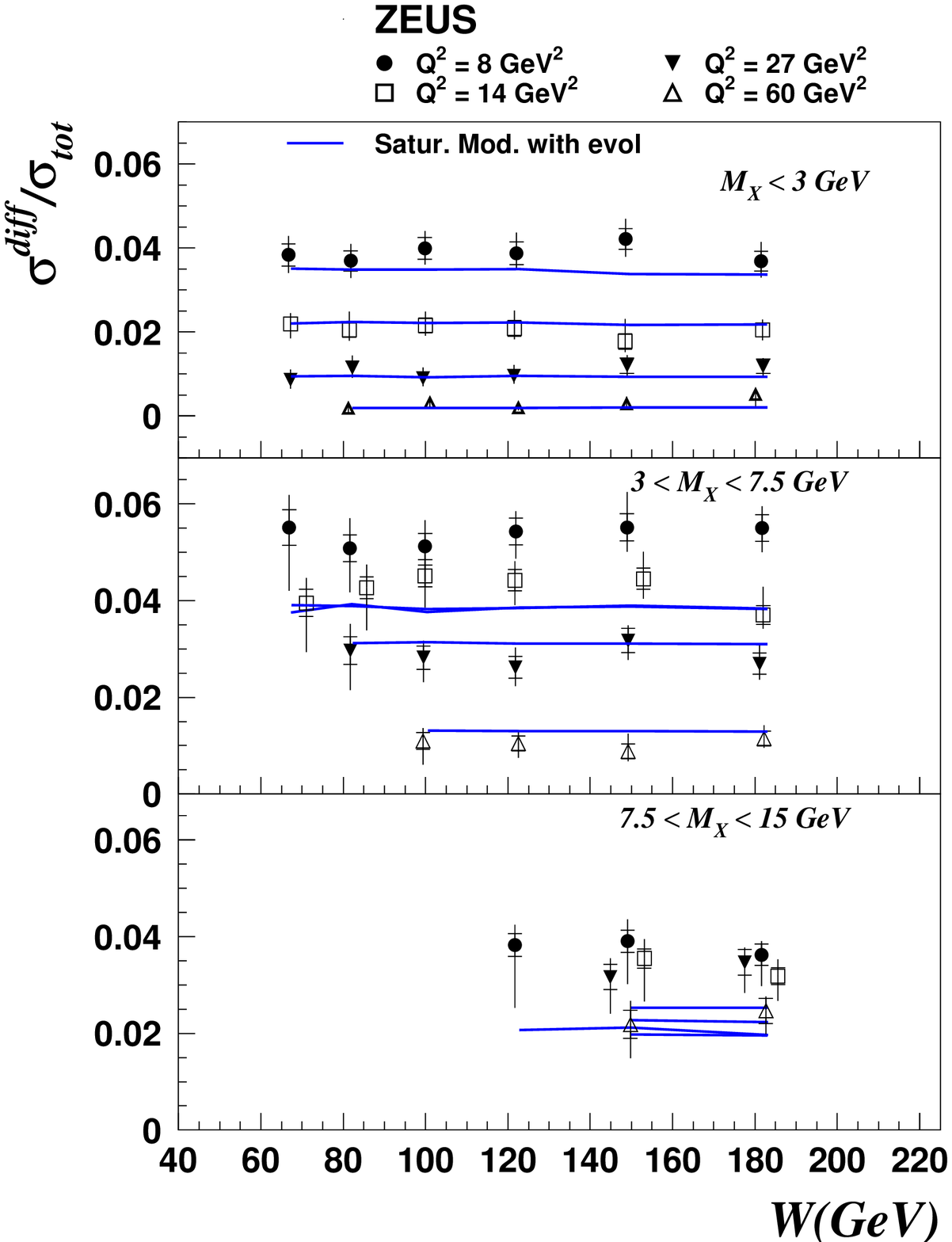}
    \hskip 3mm
    \includegraphics[width=0.35\textwidth]{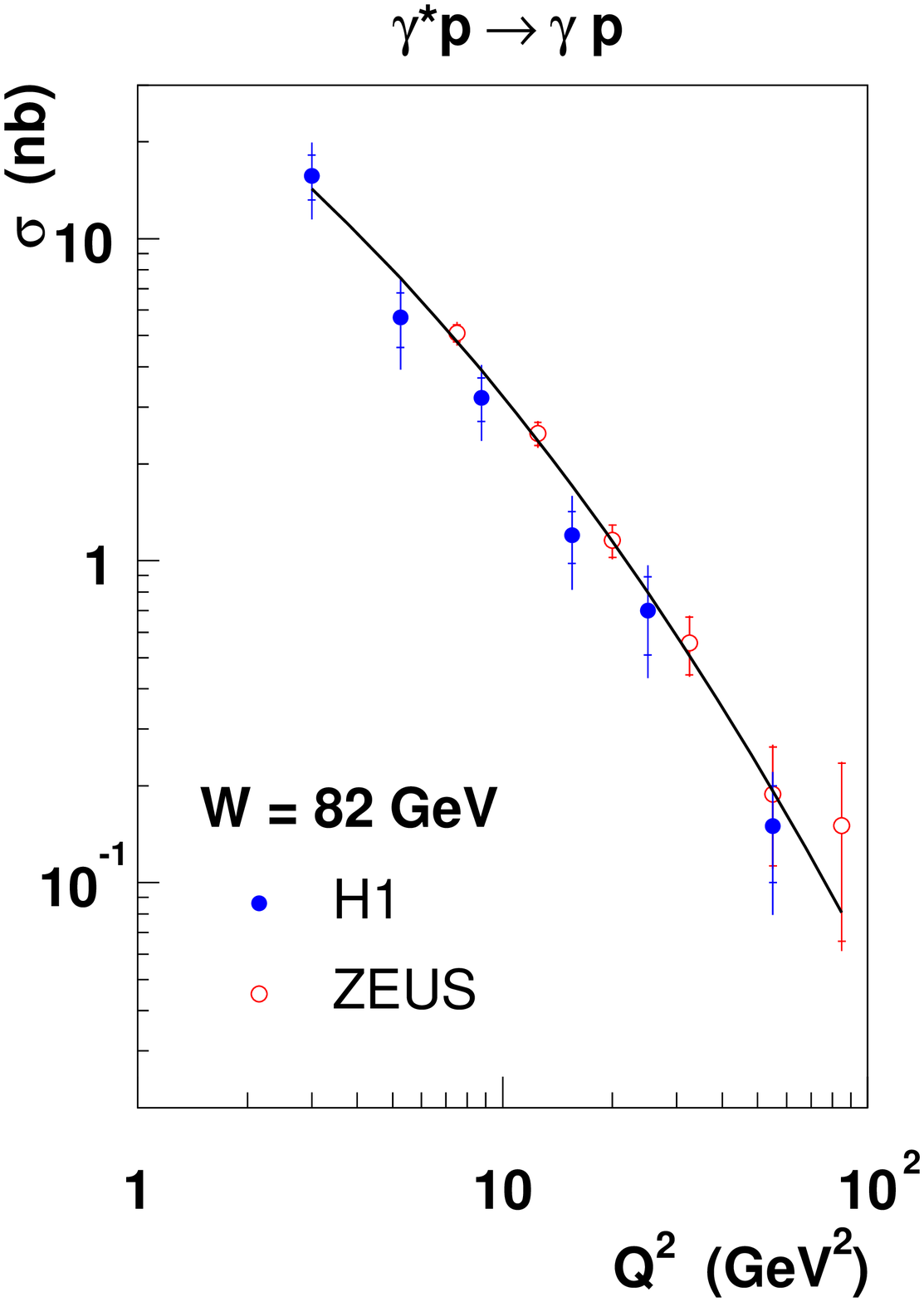} 
  \end{center}
  \caption{\baselineskip 15pt  \label{fig:diff}\sl Left: ZEUS data for the ratio
    $\sigma_{\rm diff}/\sigma_{\rm tot}$ together with the
    corresponding prediction of the saturation model in
    \cite{BarteGK1}. Right: DVCS measurement at HERA, and
    comparison with \cite{KowalMW1}.}
\end{figure*}
For small invariant masses, this process corresponds to elastic
scattering of the $q\bar q$ dipole off the target. Its cross--section
is evaluated as\footnote{Forward diffraction (corresponding to $t=0$,
  where $t=(P-P^\prime)^2$ is the momentum transfer squared between
  the incoming and outgoing proton) can be compared directly to
  inclusive DIS within a dipole model since it depends only on the
  dipole cross-section.}
 \begin{equation}
\label{sigmadiff}
\left.\frac{\rmd\sigma_{\rm diff}}{\rmd t}\right|_{t=0}
=\frac{1}{16\pi}\int\limits_0^1 \rmd z\int \rmd^2\r_\perp
\left|\psi(z,\r_\perp)\right|^2
\sigma_{\rm dipole}^2(x,\r_\perp)\; .
\end{equation}
The dipole cross-section in this expression, for small dipoles $r^2
\ll 1/Q_s^2$, is a color singlet combination of two gluons that can be
interpreted as Pomeron exchange~\cite{pomeron}; for larger dipoles
$r^2\geq 1/Q_s^2$, the color singlet exchange does not have this
simple interpretation. The $t$ distribution has the form $d\sigma_{\rm
  diff}/dt = \exp(-B_{_D} |t|) d\sigma_{\rm diff}/dt|_{t=0}$, where
$B_{_D}$ is the transverse area of the interaction region in the
proton and is closely related to the transverse gluon radius in the
proton estimated to be $0.61\pm 0.04$~fm~\cite{CaldwK1}. From this
form of the diffractive cross-section and eq.~(\ref{sigmadiff}), the
total diffractive cross-section is
\begin{eqnarray}\label{intdiff}
 \sigma_{_D} \sim
 \frac{B_{_D}}{Q^2}\int\limits_{Q^{-2}}^{Q_s^{-2}}
 \frac{\rmd r^2 }{r^4}\, \Big(r^2Q_s^2(x)\Big)^2
  \sim B_{_D}\,\frac{Q_s^2(x)}{Q^2}\; .
\end{eqnarray}
We used here the color transparency approximation $\sigma_{\rm
  dipole}\propto r^2 Q_s^2(x)$ for the dipole cross-section. Unlike
the inclusive cross-section in eq.~(\ref{eq:X-DIS}) which is dominated
by small dipole sizes $\sim 1/Q$, the integrand of the diffractive
cross-section is dominated by larger size dipoles of size $\sim
1/Q_s$.  Comparing eq.~(\ref{intdiff}) with eq.~(\ref{eq:X-DIS}), we
deduce that in the saturation framework the ratio $\sigma_{\rm
  diff}/\sigma_{\rm tot}$ is approximately constant as a function of
energy.  As shown in figure~\ref{fig:diff} (left), the HERA data
support this qualitative observation and are in quantitative agreement
with the detailed saturation model of~\cite{BarteGK1}. We note further
that excellent fits with $\chi^2 \sim 1$ are obtained in this
saturation framework for exclusive vector meson production and deeply
virtual Compton scattering~\cite{KowalMW1} in addition to inclusive
diffraction~\cite{KowalLMV1} (see figure \ref{fig:diff}, right). These
exclusive processes provide detailed information about the impact
parameter dependence of the dipole cross-section~\cite{impact} and may
even provide unique information about the partonic nature of short
range nuclear forces~\cite{CaldwK1}.

In contrast, if unitarization were due to soft physics (the prevailing
viewpoint before the advent of saturation~\cite{Bjork6}), the
natural cutoff for the integral would be $1/\Lam$. Diffraction would
be non--perturbative even for hard $Q$ and the energy dependence of
the diffractive cross-section would be the square of the inclusive
cross-section, in disagreement with data.

Fits based on geometrical scaling of the e+A fixed target data with
the functional form $Q_s^2 \propto A^\delta$ (with $\delta$ naively
$1/3$) give values of $\delta$ with range from $1/4$ to
$4/9$~\cite{delta}. However, the dipole formalism which describes e+p
data so successfully can be straightforwardly generalized to nuclei to
construct the corresponding nuclear dipole
cross-section~\cite{KowalT1}. When compared to results for the $x,Q^2$
dependence of inclusive e+A data, one finds effectively that
$Q_{s,A}^2 (x)= Q_{s,p}^2 (x) A^{1/3}$~\cite{KowalLV1}.  Nuclear
diffractive distributions can be computed in saturation
models~\cite{KowalLMV1,KugerGN1}; predictions have been made as well
for semi-inclusive hadron production~\cite{MarquXY1}, exclusive vector
meson production~\cite{vector} and nuclear DVCS~\cite{Macha1} in this
framework.

In saturation models of nuclei, the small $x$ distributions in a
nucleon are convolved with nuclear geometry to give the nuclear
distributions. This process however does not commute with the RG
evolution in $x$ of nuclear distributions determined at some initial
scale. First computations of e+A inclusive distributions in the NLO BK
framework have been performed and good agreement obtained for existing
fixed target data~\cite{DusliGLV1}.

\subsection{Particle multiplicities in d+A and A+A collisions}

The CGC EFT is most reliable when at least one of the projectiles is
dense in the sense discussed in the previous section. At RHIC, the
world's first deu\-te\-ron+hea\-vy nucleus (d+A) and A+A collider, many
features of the CGC are being tested. These include bulk features such
as the rapidity and centrality dependence of particle multiplicities
in d+A and A+A collisions, limiting fragmentation, particle spectra
and correlations, and even possibly more exotic features such as long
range rapidity correlations (``the ridge'') in A+A collisions and
local CP violation arising from topological fluctuations in A+A
collisions. In this sub-section, we will focus on particle
multiplicities and discuss the other features in subsequent
sub-sections.

``Limiting fragmentation'' is the well known property of the strong
interactions that the rapidity distribution in the fragmentation
region becomes independent of the collision energy.  When one
increases the beam energy (see figure \ref{fig:geom}, right),
$dN/d\eta^\prime$ is the same at large $\eta^\prime$ for all energies,
where $\eta^\prime\equiv \eta-\eta_{\rm beam}$.  In the fragmentation
region, large $x_1$ modes are probed in one hadron or nucleus and
small $x_2$ modes in the other. At small $x_2$, if gluonic matter is
saturated, parton distributions have a very weak dependence on $x_2$,
or equivalently on $\eta+\eta_{\rm beam}$; cross-sections will only
depend on $x_1$ or $\eta-\eta_{\rm beam}$. Deviations from limiting
fragmentation at high energies are especially interesting because of a
significant window in phase space where the RG evolution in $x_2$ to
(or away) from the universal ``black disk'' of saturated gluon matter
can be explored~\cite{Jalil3,GelisSV1}. These constraints lead to
specific predictions for the rapidity dependence of the multiplicity
at the LHC in the BK RG framework~\cite{GelisSV1,Albac1}.

The CGC can only predict the distribution of initial gluons, at a
proper time $\tau\sim Q_s^{-1}$.  In p/d+A collisions this is not a
significant limitation because only a few gluons are produced in the
final state. The situation is vastly different in A+A collisions where
the particle multiplicity is significantly larger and the system
evolves from the non-equilibrium Glasma to a QGP, the latter evolving
subsequently as a fluid with low viscosity. For p+A collisions, the
problem is solvable analytically~\cite{KovchM3,DumitM1,BlaizGV1}. For
A+A collisions, two kinds of calculations have been performed:

(a) Exact numerical solutions of the Yang-Mills
equations~\cite{class0,KrasnV2,Lappi}. Quantum evolution effects are
not included systematically and the ensemble of color charges is
assumed to be Gaussian (MV model), which is reasonable at RHIC
energies of $x\sim 10^{-2}$ at RHIC. The inclusion of quantum effects
from the wavefunctions~\cite{factorization,GelisLM1} is under control,
but is still an unsolved problem for those present in the final state
evolution~\cite{insta1,FukusGM1}.

(b) Approximate analytical calculations of the initial gluon spectrum
\cite{KharzL1}. These calculations assume $k_\perp$-factorization
although it is violated for momenta $p_\perp \leq Q_s$.  In this
model, saturation effects are introduced via the unintegrated gluon
distribution of the nuclei with the rapidity dependence of the gluon
spectrum governed by that of the saturation scale $Q_s^2(x)\sim
x^{-0.3}$. In the CGC framework, $dN/d\eta = Q_s^2
R_A^2/\alpha_s(Q_s)$; a unique feature is that the coupling runs as a
function of $Q_s$~\cite{Muell12}, an observation which is in agreement
with the centrality dependence of RHIC data~\cite{KharzN1}.  Similar
analyzes of multiplicity distributions were extended to the case of
d+A collisions \cite{KharzLN2}, with results in fair agreement with
RHIC data.

\subsection{Inclusive spectra in d+A and A+A collisions}
\subsubsection{$p_\perp$ dependence}
One does not expect significant final state interactions with on-shell
partons in $p+A$ collisions.  Moreover, the inclusive gluon spectrum
depends only on the Wilson line correlator
$\big<U(0)U^\dagger(\x_\perp)\big>$~\cite{DumitM1,pA,BlaizGV1}, that
can be determined from the dipole cross-section used in
DIS~\cite{GelisJ3}.  The hadron spectrum in d+A collisions has been
evaluated in this approach in \cite{DumitHJ1}, with results in good
agreement with RHIC data. Very recently, good agreement of single
inclusive distributions in d+A collisions were obtained in the NLO RG
approach~\cite{AlbacM1} consistent with its application in e+A
collisions~\cite{DusliGLV1}.

In A+A collisions, the strong final state interactions likely alter
the momentum distribution to be more isotropic at low momenta; the
hard tail is modified by parton energy loss. While these effects may
in part be included in the Glasma, the additional contributions at
later stages are important.

\subsubsection{Nuclear modification ratios}
To quantify nuclear effects, $p_\perp$ spectra in A+A and d+A
can be compared to those in p+p collisions by the ratios
\begin{eqnarray}
R_{AA}\equiv
\frac{\left.\frac{dN}{dyd^2\p_\perp}\right|_{AA}}
{N_{\rm coll}\left.\frac{dN}{dyd^2\p_\perp}\right|_{pp}}
\; ,\quad
R_{dA}\equiv
\frac{\left.\frac{dN}{dyd^2\p_\perp}\right|_{dA}}
{N_{\rm coll}\left.\frac{dN}{dyd^2\p_\perp}\right|_{pp}}\; ,
\end{eqnarray}
where $N_{\rm coll}\sim A^{4/3}$ (resp. $A$) is the number of binary
collisions in A+A (resp. d+A) collisions. At high $p_\perp$, one
expects the ratios to scale with $N_{\rm coll}$.  In A+A collisions at
RHIC, because no suppression is seen in d+A collisions at central
rapidities, the large observed suppression in $R_{AA}$ is rightly
interpreted as a final state effect, with a strong candidate being
energy loss induced by the dense medium~\cite{Eloss}.

However, at small $x$, RG evolution {\it a la} BK
predicts~\cite{forward} that this ratio should be suppressed by
saturation effects in the wavefunction of the nucleus.  This can be
studied in the clean environment of d+A collisions where small $x$ is
probed measuring the ratio at a positive rapidity in the direction of
the deuterium nucleus. The ratio $R_{dA}$ was measured by the
BRAHMS~\cite{Arsena1} collaboration up to rapidities of $\eta=3.2$ and
by the STAR collaboration~\cite{Adamsa6} at $\eta=4$. At these large
$\eta$, the value of $R_{dA}$ is significantly below unity, consistent
with predicted trend. In addition, as anticipated, the suppression is
greater for more central collisions. The onset of this forward
suppression was also studied semi-analytically in
\cite{BlaizGV1,IancuIT2}, and more quantitative calculations have been
performed in \cite{KKT,DumitHJ1,GoncaKMN1}, using models
of the dipole cross-section that have a realistic $x$-dependence and
most recently in the NLO BK framework~\cite{AlbacM1}.

\subsection{Two hadron correlations in d+A and A+A collisions}

\begin{figure}[htb!]
\begin{center}
\resizebox*{7.2cm}{!}{\includegraphics{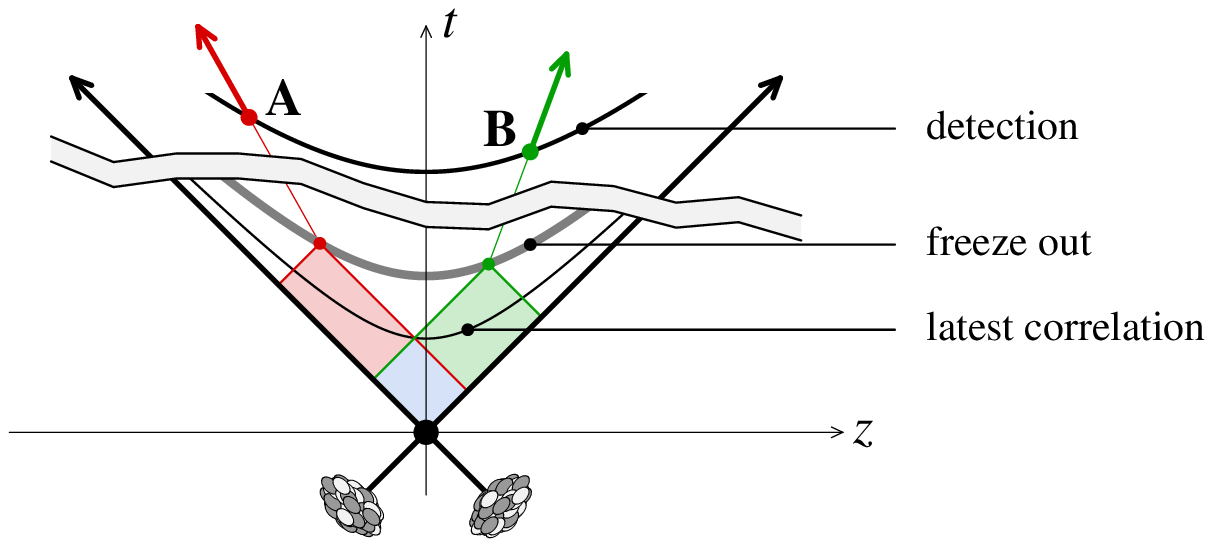}}
\resizebox*{4.8cm}{!}{\includegraphics{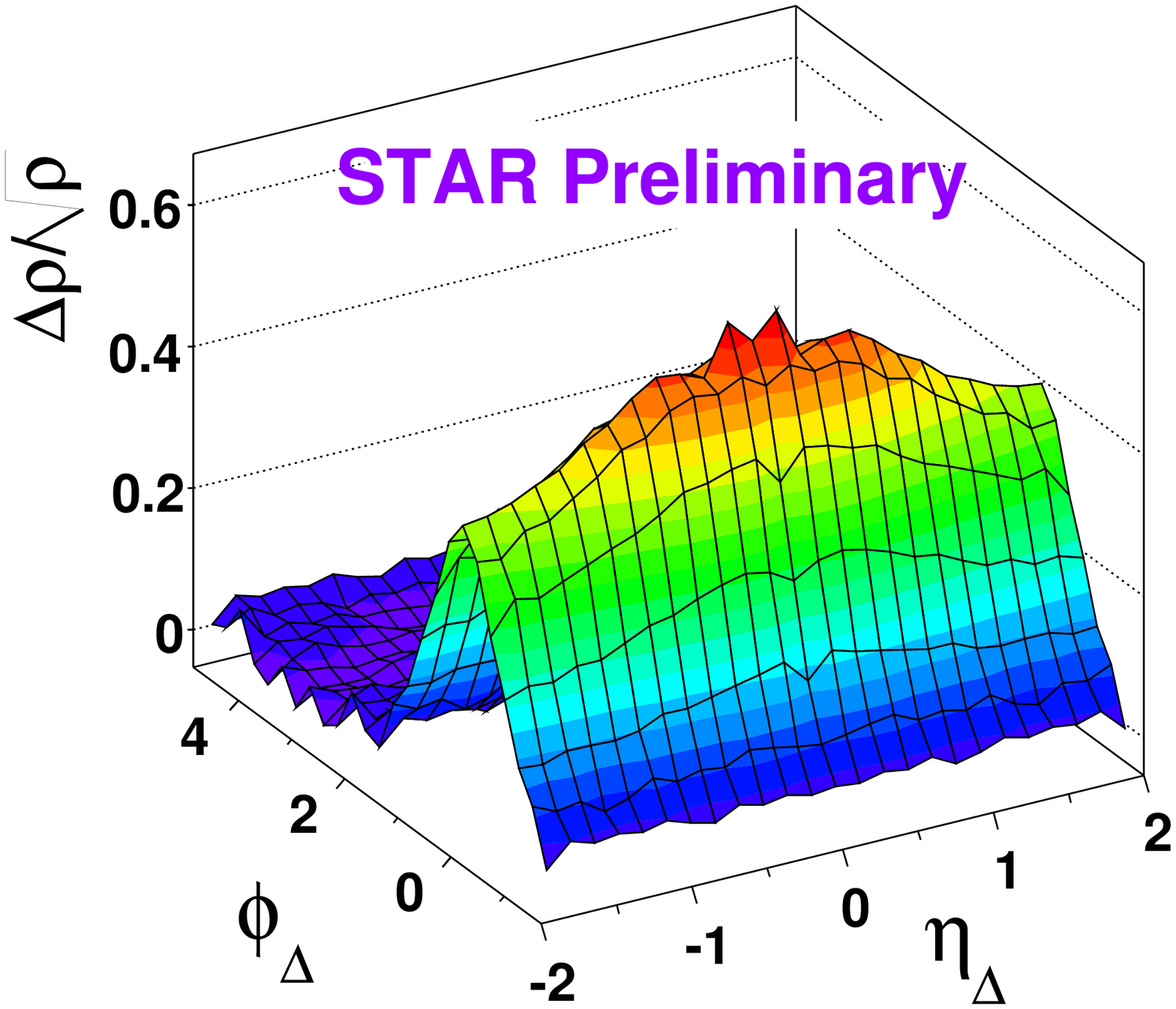}}
\end{center}
\caption{\baselineskip 15pt \label{fig:2hadron}Left: causal relations
  between two particles separated in rapidity. Right: 2-hadron
  correlation measured by STAR as a function of $\Delta\eta$ and
  $\Delta\phi$. }
\end{figure}
 
Two hadron correlations\footnote{Hadron-photon correlations have also
  been studied in \cite{Jalil6}.} are more sensitive in distinguishing
between model predictions than the single inclusive results for which
alternative explanations of the data are feasible.  Very recently,
striking preliminary results on these correlations in d+Au collisions
have been presented by the STAR and PHENIX collaborations. In the STAR
measurements, back-to-back correlations of pairs have been studied
where i) one particle in the pair is forward in rapidity and the other
at central rapidity and ii) both particles in the pair are at forward
rapidities.  The ``forward-central'' (FC) pairs probe $x\sim 10^{-2}$
in the gold nucleus while the``forward-forward'' (FF) pairs probe
$x\sim 10^{-3}$ in the gold nucleus.  A clear broadening of the
backward hadron peak is seen in the transition from FC to FF, as well
as with increasing centrality in the FF events--in the latter case,
the distribution is so broad that no peak is visible! This particular
effect was predicted in the CGC~\cite{Marqu1} and is in quantitative
agreement the STAR results~\cite{decorr}. These results are also
consistent with measurements by the PHENIX collaboration on pair
distributions in d+Au collisions. Related discussions can be found
in~\cite{pair}.

While two particle correlations in A+A collisions are typically much
altered by final state interactions, there is an important exception
for particles widely separated in rapidity. Causality implies that
these correlations are created at very early times, as illustrated in
the left panel of figure \ref{fig:2hadron}. A simple estimate for
ultra-relativistic particles whose space-time and momentum rapidities
are strongly correlated gives
\begin{equation}
\tau_{\rm max}=\tau_{\rm freezeout}\; e^{-\frac{|\Delta\eta|}{2}}\; ,
\end{equation}
for the latest time at which these particles could have been
correlated. For a freeze-out time $\tau_{\rm freezeout}\approx
10~$fm/c, and rapidity separations $\Delta\eta\geq 4$, one sees that
these correlations must have been generated before $1~$fm/c. Thus long
range rapidity correlations are a ``chronometer'' of the evolution of
the very strong Glasma color fields produced early in the heavy ion
collision.

Striking long range rapidity correlations were seen in A+A collisions
in events with prominent jet like structures, where the spectrum of
associated particles is observed to be collimated in the azimuthal
separation $\Delta \varphi$ relative to the jet and shows a nearly
constant amplitude in the strength of the pseudo-rapidity correlation
$\Delta \eta$ up to $\Delta \eta\sim 1.5$~\cite{Adamsa4}. These events
were coined ``ridge'' events following from the visual appearance of
these structures as an extended mountain ridge in the $\Delta
\eta$-$\Delta \varphi$ plane associated with a narrow jet peak.  This
collimated correlation persists up to $\Delta \eta ~\sim
4$~\cite{Molna1,Alvera1}. An important feature of ridge correlations
is that the above described structure is seen in two particle
correlations without a jet trigger (see figure~\ref{fig:2hadron}) and
persists without significant modification for the triggered
events~\cite{Adamsa5,Abelea1}.  These events include all hadrons with
momenta $p_\perp \geq 150$ MeV. In such events, a sharp rise in the
amplitude of the ridge is seen~\cite{Adamsa5,Daugh2} in going from
peripheral to central collisions.

The ridge structures seen in two particle and three particle~\cite{3p}
A+A correlations can be explained as resulting from the transverse
radial flow of the Glasma flux tubes we discussed previously in
section \ref{sec:AA}. The flux tubes are responsible for the long
range rapidity correlation in the ridge; the angular collimation
occurs because particles produced isotropically in a given flux tube
are collimated by the radial outward ``Hubble'' hydrodynamic flow of
the flux tubes. Ideas on the angular collimation of particle
distributions by flow were discussed in refs.~\cite{flow}. When
combined with the long range rapidity correlations provided by the
Glasma flux tubes, they provide a semi-quantitative description of the
ridge measurements~\cite{DumitGMV1,gavin,DusliGLV1,DusliFV1}.  More
detailed studies are feasible with realistic hydrodynamic
simulations~\cite{hydro}.  We note that the PHENIX collaboration which
also observed ridge-like structures~\cite{Adarea3} have presented
preliminary data showing that ridge-like structures disappear as
the $p_\perp$ of the trigger particle is increased~\cite{Chen1}; this
result is in qualitative agreement with expectations in the Glasma
flux tube picture. For a recent critical evaluation of this and
alternative models, we refer the reader to ~\cite{Nagle1}.

\section{Outlook: LHC and future DIS colliders}

At the LHC, very low values of $x$ will be probed in p+p, A+A
(including the photo-production induced dynamics of peripheral A+A
collisions) and possibly p+A collisions through a wide range of final
states and diverse kinematic ranges. In nuclei, $Q_s^2$ will be large;
estimates range from $\sim 2.6$-$4$ GeV$^2$ in central collisions to
$\sim 10$ GeV$^2$ for $y=\pm 3$ units. The picture of strongly
correlated albeit weakly coupled dynamics of glue in the CGC EFT
outlined here will be tested as never before. Detailed tests of BFKL
dynamics and possibly even the CGC are feasible in p+p collisions in
studies respectively of Mueller-Navelet~\cite{MuellN1} and forward
jets.  Diffractive final states, while always challenging to
interpret, offer opportunities to understand deeply how saturated
gluons generate rapidity gaps. A strong test of CGC dynamics will
already be available in ``Day 1'' physics of bulk dynamics in A+A
collisions. More subtle tests of the RG flow of multi-parton
correlators are available in a variety of final states ranging from
quarkonia to electromagnetic probes to long range rapidity
correlations. In p+A collisions at the LHC, many of the patterns seen
in forward single inclusive and di-hadron correlation spectra will be
present already at central rapidities and will be much more dramatic
at forward rapidities. Finally, the role of Glasma dynamics
relative to that of the Quark-Gluon Plasma in A+A collisions at the
LHC is still unclear. If RHIC has reached the perfect hydrodynamic
limit of maximal flow, what happens at the LHC?

Studies of the dynamics of gluon saturation and the CGC that are
complementary to the LHC can be performed at a future Electron-Ion
Collider (EIC)\footnote{Current proposals include the EIC in the eRHIC
  version at BNL and the eLIC version at Jlab, the LHeC proposal at
  CERN and an electron-ion collider at FAIR in GSI.}. Firstly, with a
wide lever arm in $Q^2$, the dynamics of saturation can be studied
with precision in the regime where $Q^2\sim Q_s^2\gg \Lambda_{_{\rm
    QCD}}^2$; this is difficult to achieve at a hadron collider.
Secondly, clarifying what aspects of the dynamics probed are universal
and novel calls for an electron probe. The factorization theorems we
have developed here suggest that the density matrices $W$ are
universal--how can we confirm this and cleanly extract their rich
dynamics? These issues are not merely academic and are strongly
reflected in the structure of final states.  For example, rapidity
gaps are a large fraction of the cross-section at HERA but are a much
smaller contribution in hadronic collisions at Fermilab, demonstrating
a breakdown of factorization~\cite{AlverCTW1}.  The physics case for
an EIC/LHeC has been outlined in \cite{EIC} and in a
number of white papers and reports.

In summary, the prospects for unambiguous discovery and exploration of
a novel many body QCD regime of gluon saturation are very bright in
the next decade. The theoretical status of studies of this regime
within the framework of the CGC EFT are increasingly robust albeit may
questions remain. It is hoped that experiments, as usual, will provide
definitive answers.

\section*{Acknowledgements}
F.G. and E.I. are supported in part by Agence Nationale de la
Recherche via the program ANR-06-BLAN-0285-01. J.J.-M. is supported by
the DOE Office of Nuclear Physics through Grant No.~DE-FG02-09ER41620
and by the City University of New York through the PSC-CUNY Research
Award Program, grant 62625-40. R.V.'s research is supported
by the US Department of Energy under DOE Contract No.
DE-AC02-98CH10886.

%\bibliography{biblio}
%\bibliographystyle{unsrt}

\end{document}